\documentclass{amsart}
\usepackage{amsmath}
\usepackage{latexsym}
\usepackage{amsfonts}
\usepackage{amssymb}
\usepackage{graphicx}
\newtheorem{theorem}{Theorem}

\newtheorem{lemma}[theorem]{Lemma}

\newtheorem{proposition}[theorem]{Proposition}

\begin{document}

\author[Hall]{Brian C. Hall}
\address{Department of Mathematics\\
University of Notre Dame \\
Notre Dame, IN 46556, USA }
\email{bhall{@}nd.edu}
\author[Mitchell]{Jeffrey J. Mitchell}
\address{Department of Mathematics\\
Baylor University \\
Waco, TX 76798, USA }
\email{jeffrey\_mitchell{@}baylor.edu}
\title{Coherent states on spheres}
\begin{abstract}
We describe a family of coherent states and an associated resolution of the
identity for a quantum particle whose classical configuration space is the $%
d $-dimensional sphere $S^{d}.$ The coherent states are labeled by points in
the associated phase space $T^{\ast }(S^{d}).$ These coherent states are 
\textit{not} of Perelomov type but rather are constructed as the
eigenvectors of suitably defined annihilation operators.

We describe as well the Segal--Bargmann representation for the system, the
associated unitary Segal--Bargmann transform, and a natural inversion
formula. Although many of these results are in principle special cases of
the results of B. Hall and M. Stenzel, we give here a substantially
different description based on ideas of T. Thiemann and of K. Kowalski and
J. Rembieli\'{n}ski.

All of these results can be generalized to a system whose configuration
space is an arbitrary compact symmetric space. We focus on the sphere case
in order to carry out the calculations in a self-contained and explicit way.
\end{abstract}
\maketitle
\tableofcontents

\section{Introduction\label{intro.sec}}

In \cite{H1} B. Hall introduced a family of coherent states for a system whose
classical configuration space is the group manifold of a compact Lie group
$G.$ These coherent states are labeled by points in the associated phase
space, namely the cotangent bundle $T^{\ast}(G).$ The coherent states
themselves were originally defined in terms of the \textit{heat kernel} on
$G,$ although we will give a different perspective here. One may identify
\cite{H3} $T^{\ast}(G)$ with the complexified group $G_{\mathbb{C}}$, where,
for example, if $G=\mathrm{SU}(2)$ then $G_{\mathbb{C}}=\mathrm{SL}%
(2;\mathbb{C)}$. The paper \cite{H1} establishes a resolution of the identity
for these coherent states, and equivalently, a unitary Segal--Bargmann
transform. The Segal--Bargmann representation of this system is a certain
Hilbert space of holomorphic functions over the complex group $G_{\mathbb{C}%
}.$ Additional results may be found in \cite{H3,H2,H4} and the survey paper
\cite{H7}.

The coherent states for $G$ (in the form of the associated Segal--Bargmann
transform) have been applied to quantum gravity in \cite{A}, with proposed
generalizations due to Thiemann \cite{T1}. More recently the coherent states
themselves have been used by Thiemann and co-authors \cite{T2} in an attempt
to determine the classical limit of the quantum gravity theory proposed by
Thiemann in \cite{QSD}. In particular, the second entry in \cite{T2}
establishes good phase space localization properties (in several different
senses) for the coherent states associated to the configuration space
$G=\mathrm{SU}(2).$

In another direction, K. K. Wren \cite{Wr}, using a method proposed by N. P.
Landsman \cite{L1}, has shown that the coherent states for $G$ arise naturally
in the canonical quantization of $(1+1)$-dimensional Yang-Mills theory on a
spacetime cylinder. Here $G$ is the structure group of the theory and plays
the role of the reduced configuration space, that is, the space of connections
modulo based gauge transformation over the spatial circle. Wren considers
first the ordinary canonical coherent states for the unreduced
(infinite-dimensional) system. He then shows that after ``projecting'' them
into the gauge-invariant subspace (using a suitable regularization procedure)
these become precisely the generalized coherent states for $G,$ as originally
defined in \cite{H1}. Driver and Hall \cite{DH} elaborate on the results of
Wren, using a different regularization scheme. They show in particular how the
resolution of the identity for the generalized coherent states can be obtained
by projection from the resolution of the identity for the canonical coherent
states. See also \cite{AHS} for an appearance of the generalized
Segal--Bargmann transform in the setting of 2-dimensional Euclidean Yang-Mills theory.

Finally, the paper \cite{H9} shows that the generalized Segal--Bargmann
transform for $G$ can be obtained by means of geometric quantization (see also
\cite[Sect. 3.2]{H7}). This means that the associated coherent states for $G$
are of ``Rawnsley type'' \cite{Ra} and are thus in the spirit of Berezin's
approach to quantization.

We emphasize that the coherent states for $G$ are \textit{not of Perelomov
type} \cite{P}. Instead they are realized as the eigenvectors of certain
non-self-adjoint ``annihilation operators,'' as will be described in detail in
the present paper. (See Section \ref{conclusions.sec} for further comments.)

The coherent states and the resolution of the identity for $G$ ``descend'' is
a straightforward way to the case of a system whose configuration space is a
compact symmetric space $X$ \cite[Sect. 11]{H1}. Compact symmetric spaces are
manifolds of the form $G/K,$ where $G$ is a compact Lie group and $K$ is a
special sort of subgroup, namely, the fixed-point subgroup of an involution.
Examples include the spheres $S^{d}=\mathrm{SO}(d+1)\mathrm{/SO}(d)$ and the
complex projective spaces $\mathbb{C}P^{d}=\mathrm{SU}(d+1)/\mathrm{SU}(d).$
Compact Lie groups themselves can be thought of as symmetric spaces by
identifying $G$ with $\left(  G\times G\right)  /\Delta(G) $ where $\Delta(G)$
is the ``diagonal'' copy of $G$ inside $G\times G.$

We emphasize that in the case $X=S^{2}$ the 2-sphere is playing the role of
the \textit{configuration space} and thus the coherent states discussed here
are completely different from the spin coherent states in which the 2-sphere
plays the role of the phase space. Whereas the spin coherent states are
labeled by points in $S^{2}$ itself, our coherent states are labeled by points
in the cotangent bundle $T^{\ast}(S^{2}).$

Although the case of compact symmetric spaces can be treated by descent from
the group, it is preferable to give a direct treatment, and such a treatment
was given by Stenzel \cite{St}. In particular Stenzel gives a much better
description, in the symmetric space case, of the measure that one uses to
construct the resolution of the identity. (See also \cite[Sect. 3.4]{H7}.)
Although Stenzel formulates things in terms of a unitary Segal--Bargmann
transform and does not explicitly mention the coherent states, only a
notational change is needed to re-express his results as a resolution of the
identity for the associated coherent states.

More recently, the coherent states for the 2-sphere $S^{2}$ were independently
discovered, from a substantially different point of view, by Kowalski and
Rembieli\'{n}ski \cite{KR1}. (The paper \cite{KR1} builds on earlier work of
Kowalski, Rembieli\'{n}ski, and Papaloucas \cite{KRP} on the $S^{1}$ case.)
These authors were unaware at the time of the work of Hall and Stenzel. The
paper \cite{KR2} then describes the resolution of the identity for the
coherent states on the 2-sphere, showing in a different and more explicit way
that the result of \cite[Thm. 3]{St} holds in this case. (See Section VII of
\cite{KR2} for comments on the relation of their work to that of Stenzel.)

The purpose of this paper is to describe the coherent states for a compact
symmetric space using the points of view advocated by Thiemann and by Kowalski
and Rembieli\'{n}ski. For the sake of concreteness we concentrate in this
paper on the case $X=S^{d}.$ In \cite{H1} and \cite{St} the coherent states
are defined in terms of the heat kernel on the configuration space, which
takes the place of the Gaussian that enters into the description of the
canonical coherent states. Here by contrast the coherent states are defined to
be the eigenvectors of suitable annihilation operators, and only afterwards
does one discover the role of the heat kernel, in the position wave function
of the coherent states and in the reproducing kernel. The annihilation
operators, meanwhile, are defined by (a special case of) the ``complexifier''
method proposed by Thiemann, which we will show is equivalent to (a
generalization of) the polar-decomposition method of Kowalski and
Rembieli\'{n}ski. We emphasize, though, that the approach described in this
paper gives ultimately the same results as the heat kernel approach of Hall
and Stenzel.

\section{Main results\label{main.sec}}

In this section we briefly summarize the main results of the paper. All
results are explained in greater detail in the subsequent sections. Briefly,
our strategy is this. First, we construct complex-valued functions
$a_{1},\cdots,a_{d+1}$ on the classical phase space that serve to define a
complex structure on phase space. Second, we construct the quantum
counterparts of these functions, operators $A_{1},\cdots,A_{d+1}$ that we
regard as the annihilation operators. Third, we construct simultaneous
eigenvectors for the annihilation operators, which we regard as the coherent
states. Fourth, we construct a resolution of the identity for these coherent states.

We consider a system whose classical configuration space is the $d$%
-dimensional sphere $S^{d}$ of radius $r.$ We consider also the corresponding
phase space, the cotangent bundle $T^{\ast}(S^{d}),$ which we describe as
\[
T^{\ast}(S^{d})=\left\{  \left.  (\mathbf{x},\mathbf{p})\in\mathbb{R}%
^{d+1}\times\mathbb{R}^{d+1}\right|  x^{2}=r^{2},\,\mathbf{x}\cdot
\mathbf{p}=0\right\}  ,
\]
where $\mathbf{p}$ is the \textit{linear} momentum.

In Section \ref{complex.sec} we consider the classical component of Thiemann's
method. To apply this method we must choose a constant $\omega$ with units of
frequency. The classical ``complexifier'' is then defined to be
\textit{kinetic energy function divided by }$\omega,$ which can be expressed
as
\[
\text{complexifier}=\frac{\text{kinetic energy}}{\omega}=\frac{j^{2}}{2m\omega
r^{2}},
\]
where $j^{2}$ is the total angular momentum. (Thiemann's method allows other
complexifiers; see Section 3.) We construct complex-valued functions
$a_{1},\cdots,a_{d+1}$ by taking the position functions $x_{1},\cdots,x_{d+1}$
and applying repeated Poisson brackets with the complexifier. Specifically,
\begin{align}
a_{k}  & =e^{i\left\{  \cdot,\text{complexifier}\right\}  }x_{k}\nonumber\\
& =\sum_{n=0}^{\infty}\left(  \frac{i}{2m\omega r^{2}}\right)  ^{n}\frac
{1}{n!}\underset{n}{\underbrace{\left\{  \cdots\left\{  \left\{  x_{k}%
,j^{2}\right\}  ,j^{2}\right\}  ,\cdots,j^{2}\right\}  }}\label{a.intro}%
\end{align}
If we let $\mathbf{a=}\left(  a_{1},\cdots,a_{d+1}\right)  $ then the
calculations in Section \ref{complex.sec} will give the following explicit
formula
\begin{equation}
\mathbf{a}=\cosh\left(  \frac{j}{m\omega r^{2}}\right)  \mathbf{x}%
+i\frac{r^{2}}{j}\sinh\left(  \frac{j}{m\omega r^{2}}\right)  \mathbf{p}%
.\label{a.intro2}%
\end{equation}
These complex-valued functions satisfy $a_{1}^{2}+\cdots+a_{d+1}^{2}=r^{2}$
and $\left\{  a_{k},a_{l}\right\}  =0.$ In the case $d=2$ this agrees with Eq.
(6.1) of \cite{KR1}.

In Section \ref{annihilate.sec} we consider the quantum component of
Thiemann's method. We consider the quantum counterpart of the classical
complexifier, namely,
\[
\text{complexifier}=\frac{\text{kinetic energy}}{\omega}=\frac{J^{2}}{2m\omega
r^{2}},
\]
where $J^{2}$ is the total angular momentum operator. Then if $X_{1}%
,\cdots,X_{d+1}$ denote the position operators we define, by analogy with
(\ref{a.intro}),
\begin{align}
A_{k}  & =e^{i[\cdot,\text{ complexifier}]/i\hbar}X_{k}\nonumber\\
& =\sum_{n=0}^{\infty}\left(  \frac{1}{2m\omega r^{2}\hbar}\right)  ^{n}%
\frac{1}{n!}\underset{n}{\underbrace{\left[  \cdots\left[  \left[  X_{k}%
,J^{2}\right]  ,J^{2}\right]  ,\cdots,J^{2}\right]  }}.\label{qa.intro}%
\end{align}
This may also be written as
\begin{equation}
A_{k}=e^{-J^{2}/2m\omega r^{2}\hbar}X_{k}e^{J^{2}/2m\omega r^{2}\hbar
}.\label{qa.intro2}%
\end{equation}
Equation (30) in Section \ref{annihilate.sec} gives the quantum counterpart of
(\ref{a.intro2}); it is slightly more complicated than (\ref{a.intro2})
because of quantum corrections. The annihilation operators satisfy $A_{1}%
^{2}+\cdots+A_{d+1}^{2}=r^{2}$ and $\left[  A_{k},A_{l}\right]  =0.$ Applying
the same procedure in the $\mathbb{R}^{d}$ case produces the usual complex
coordinates on phase space and the usual annihilation operators (Section
\ref{rd.sec}).

One can easily deduce from (\ref{qa.intro2}) a ``polar decomposition'' for the
annihilation operator, given in (\ref{qpolar}) in Section \ref{annihilate.sec}%
. In the case $d=2$ this is essentially the same as what Kowalski and
Rembieli\'{n}ski take as the definition of the annihilation operators. This
shows that Thiemann's complexifier approach is equivalent to the polar
decomposition approach of Kowalski and Rembieli\'{n}ski. Similarly, the polar
form of the annihilation operators in the $d=1$ case is essentially the same
as what Kowalski, Rembieli\'{n}ski, and Papaloucas take as the definition of
the annihilation operator in \cite{KRP}.

In Section \ref{coherent.sec} we consider the coherent states, defined to be
the simultaneous eigenvectors of the annihilation operators. Using
(\ref{qa.intro2}) we may immediately write down some eigenvectors for the
$A_{k}$'s, namely, the vectors of the form
\begin{equation}
\left|  \psi_{\mathbf{a}}\right\rangle =e^{-J^{2}/2m\omega r^{2}\hbar}\left|
\delta_{\mathbf{a}}\right\rangle ,\label{formula}%
\end{equation}
where $\left|  \delta_{\mathbf{a}}\right\rangle $ is a simultaneous
eigenvector for the position operators corresponding to a point $\mathbf{a}$
in $S^{d}.$ A key result of Section \ref{coherent.sec} is that one can perform
an analytic continuation with respect to the parameter $\mathbf{a},$ thereby
obtaining coherent states $\left|  \psi_{\mathbf{a}}\right\rangle $
corresponding to any point $\mathbf{a}$ in the \textit{complexified} sphere,
$S_{\mathbb{C}}^{d}=\left\{  \mathbf{a}\in\mathbb{C}^{d+1}|\,a^{2}%
=r^{2}\right\}  .$ These vectors $\left|  \psi_{\mathbf{a}}\right\rangle $ are
normalizable and satisfy
\[
A_{k}\left|  \psi_{\mathbf{a}}\right\rangle =a_{k}\left|  \psi_{\mathbf{a}%
}\right\rangle ,\quad\mathbf{a}\in S_{\mathbb{C}}^{d}.
\]

Equation (\ref{formula}) shows that the coherent states are expressible in
terms of the heat kernel on the sphere, thus demonstrating that Thiemann's
definition of the coherent states is equivalent to the definition in
\cite{H1,St} in terms of the heat kernel. The reproducing kernel for these
coherent states is also expressed in terms of the heat kernel on the sphere.

In Section \ref{res.sec} we describe a resolution of the identity for these
coherent states. In a suitable coordinate system this takes the form
\begin{equation}
I=\int_{\mathbf{x}\in S^{d}}\int_{\mathbf{p}\cdot\mathbf{x}=0}\left|
\psi_{\mathbf{a}}\right\rangle \left\langle \psi_{\mathbf{a}}\right|
\,\nu\left(  2\tau,2p\right)  \left(  \frac{\sinh2p}{2p}\right)  ^{d-1}%
\,2^{d}d\mathbf{p}\,d\mathbf{x}\label{intro.res}%
\end{equation}
where $\mathbf{a}$ is a function of $\mathbf{x}$ and $\mathbf{p}$ as in
(\ref{a.intro2}). Here $\nu$ is the heat kernel for $d$-dimensional hyperbolic
space and $\tau$ is the dimensionless quantity given by $\tau=\hbar/m\omega
r^{2}.$ Explicit formulas for $\nu$ are found in Section \ref{res.sec}. The
resolution of the identity for the coherent states is obtained by a continuous
deformation of the resolution of the identity for the position eigenvectors.

In Section \ref{sb.sec} we discuss the Segal--Bargmann representation for this
system, namely, the space of holomorphic functions on the complexified sphere
that are square-integrable with respect to the density in (\ref{intro.res}).
We think of the Segal--Bargmann representation as giving a sort of
\textit{phase space wave function} for any state. There is an inversion
formula stating the position wave function can be obtained from the phase
space wave function by integrating out the momentum variables, specifically,
\[
\left\langle \left.  \delta_{\mathbf{x}}\right|  \phi\right\rangle
=\int_{\mathbf{p}\cdot\mathbf{x}=0}\left\langle \left.  \psi_{\mathbf{a}%
(\mathbf{x},\mathbf{p})}\right|  \phi\right\rangle \nu(\tau,p)\left(
\frac{\sinh p}{p}\right)  ^{d-1}\,d\mathbf{p}%
\]
for any state $\left|  \phi\right\rangle .$ Note that whereas the resolution
of the identity involves $\nu(2\tau,2p),$ the inversion formula involves
$\nu(\tau,p)$.

In Section \ref{rd.sec} we show that the complexifier method, when applied to
the $\mathbb{R}^{d}$ case, yields the usual canonical coherent states and
their resolution of the identity. In Section \ref{euclid.sec} we summarize
some of the relevant representation theory for the Euclidean group. Finally,
in Section \ref{conclusions.sec} we compare our construction to other
constructions of coherent states on spheres.

Although all of the results here generalize to arbitrary compact symmetric
spaces $X,$ we concentrate for the sake of explicitness on the case $X=S^{d}.
$ We will describe the general case in a forthcoming paper.

\section{Complex coordinates on phase space\label{complex.sec}}

In this section we define Poisson-commuting complex-valued functions
$a_{1},\cdots,a_{d+1}$ on the classical phase space. In Section
\ref{annihilate.sec} we will introduce the quantum counterparts of these
functions, commuting non-self-adjoint operators $A_{1},\cdots,A_{d+1}$ which
we regard as the annihilation operators for this system. In Section
\ref{coherent.sec} we will consider the coherent states, that is, the
simultaneous eigenvectors of the annihilation operators.

Consider the $d$-dimensional sphere of radius $r$ in $\mathbb{R}^{d+1},$
namely,
\[
S^{d}=\left\{  \left.  \mathbf{x}\in\mathbb{R}^{d+1}\right|  x_{1}^{2}%
+\cdots+x_{d+1}^{2}=r^{2}\right\}  ,
\]
regarded as the \textit{configuration space} for a classical system ($d\geq1
$). Then consider the associated phase space, the cotangent bundle $T^{\ast
}(S^{d}),$ which we think of as
\[
T^{\ast}(S^{d})=\left\{  \left.  \left(  \mathbf{x},\mathbf{p}\right)
\in\mathbb{R}^{d+1}\times\mathbb{R}^{d+1}\right|  \,x^{2}=r^{2},\,\mathbf{x}%
\cdot\mathbf{p}=0\right\}  .
\]
Here $p$ is the \textit{linear} momentum, which must be tangent to $S^{d},$
i.e. perpendicular to $\mathbf{x}.$

We also have the \textbf{angular momentum} functions $j_{kl},$ $1\leq k,l\leq
d+1,$ given by
\begin{equation}
j_{kl}=p_{k}x_{l}-p_{l}x_{k}.\label{jkl.def}%
\end{equation}
We may think of $j$ as a function on $T^{\ast}(S^{d})$ taking values in the
space of $\left(  d+1\right)  \times\left(  d+1\right)  $ skew-symmetric
matrices, that is, in the Lie algebra $\mathrm{so}(d+1)$. Thinking of $j$ as a
matrix we may re-write (\ref{jkl.def}) as
\[
\mathbf{j}\left(  \mathbf{x},\mathbf{p}\right)  =\mathbf{p}\otimes
\mathbf{x}-\mathbf{x}\otimes\mathbf{p},
\]
where $\otimes$ denotes the outer product. (That is, $\left(  \mathbf{a}%
\otimes\mathbf{b}\right)  _{kl}=a_{k}b_{l}.$)

For a particle constrained to the sphere it is possible and convenient to
express everything in terms of $\mathbf{x}$ and $\mathbf{j}$ instead of
$\mathbf{x}$ and $\mathbf{p}.$ We may alternatively describe $T^{\ast}(S^{d})
$ as the set of pairs $\left(  \mathbf{x},\mathbf{j}\right)  $ in which
$\mathbf{x}$ is a vector in $\mathbb{R}^{d+1},$ $\mathbf{j}$ is a
$(d+1)\times(d+1)$ skew-symmetric matrix, and $\mathbf{x}$ and $\mathbf{j}$
satisfy
\begin{equation}
x^{2}=r^{2}\label{constraint.1}%
\end{equation}
and
\begin{equation}
r^{2}j_{kl}=j_{km}x_{m}x_{l}-x_{k}j_{lm}x_{m}\label{constraint.2}%
\end{equation}
(sum convention). This last condition says that if we \textit{define}
$\mathbf{p}$ to be $r^{-2}\mathbf{jx},$ then $\mathbf{j}=\mathbf{p}%
\otimes\mathbf{x}-\mathbf{x}\otimes\mathbf{p}.$ Equation (\ref{constraint.2})
reflects the constraint to the sphere and does not hold for a general particle
in $\mathbb{R}^{d+1}.$ On $T^{\ast}(S^{d})$ we have the relations
\begin{align}
\mathbf{jx}  & =r^{2}\mathbf{p}\nonumber\\
\mathbf{jp}  & =-p^{2}\mathbf{x}.\label{jj2}%
\end{align}
Recall that $\mathbf{j}$ is a matrix; thus $\mathbf{jx}$ is the vector
obtained by applying the matrix $\mathbf{j}$ to the vector $\mathbf{x},$ and
similarly for $\mathbf{jp}.$

In the case $d=2$ ($S^{2}$ sitting inside $\mathbb{R}^{3}$) a standard vector
identity shows that for any vector $\mathbf{v}\in\mathbb{R}^{3}$,
$\mathbf{jv}=(\mathbf{x}\times\mathbf{p})\times\mathbf{v},$ where $\times$ is
the cross-product and $\mathbf{x}\times\mathbf{p}$ is the usual angular
momentum vector $\mathbf{l}.$ So in the $\mathbb{R}^{3}$ case $\mathbf{jv}%
=\mathbf{l}\times\mathbf{v}.$

The symplectic structure on $T^{\ast}(S^{d})$ may be characterized by the
Poisson bracket relations
\begin{align}
\left\{  j_{kl},j_{mn}\right\}   & =\delta_{kn}j_{lm}+\delta_{lm}j_{kn}%
-\delta_{km}j_{l{}n}-\delta_{l{}n}j_{km}\nonumber\\
\left\{  x_{k},j_{lm}\right\}   & =\delta_{kl}x_{m}-\delta_{km}x_{l}%
\label{xj}\\
\left\{  x_{k},x_{l}\right\}   & =0.\nonumber
\end{align}
These are the commutation relations for the Euclidean Lie algebra, which is
the semidirect product $\mathrm{e}(d+1)\cong\mathrm{so}(d+1)\ltimes
\mathbb{R}^{d+1}.$

Poisson bracket relations involving $\mathbf{p}$ should be derived from
(\ref{xj}) using the relation $\mathbf{p}=r^{-2}\mathbf{jx}.$ Since the
constraint to the sphere alters the dynamics and hence the Poisson bracket
relations, we will not get the same formulas as in $\mathbb{R}^{d+1}.$ For
example, we have
\begin{equation}
\left\{  x_{k},p_{l}\right\}  =\delta_{kl}-\frac{x_{k}x_{l}}{r^{2}%
}.\label{xp.bracket}%
\end{equation}

The complex coordinates on phase space will be constructed from the position
functions $x_{k}$ by means of repeated Poisson brackets with a multiple of the
kinetic energy function. In the sphere case it is convenient to express the
kinetic energy in terms of the total angular momentum $j^{2}$, given by
\begin{equation}
j^{2}=\sum_{k<l}\left(  j_{kl}\right)  ^{2}.\label{j2.def}%
\end{equation}
The total angular momentum satisfies $j^{2}=r^{2}p^{2},$ and the kinetic
energy is $p^{2}/2m=j^{2}/2mr^{2}.$

We now choose a constant $\omega$ with units of a frequency. The significance
of this constant is that $m\omega$ has the units of momentum divided by
position. Thus $\omega$ (together with $m$) allows us to put position and
momentum onto the same scale, which is necessary in order define
complex-valued functions that involve both $x$ and $p.$ Ultimately, $m\omega$
will control the ratio of the width in position space of the coherent states
to the radius of the sphere.

Kowalski and Rembieli\'{n}ski do not have a parameter comparable to our
$\omega$; the only dimensional parameters in \cite{KR1} are $m,$ $r,$ and
$\hbar.$ This affects the interpretation of their Eq. (6.1) for the complex
coordinates on phase space (what we call $\mathbf{a}$). Equation (6.1)
involves $\cosh l$ and $\sinh l,$ where $l$ is the classical angular momentum.
The argument of $\cosh$ and $\sinh$ should be dimensionless, and the only way
to make $l$ dimensionless using only $m,$ $r,$ and $\hbar$ is to divide $l$ by
$\hbar.$ Thus in Eq. (6.1) of \cite{KR1} $l$ implicitly means $l/\hbar.$ In
our view it is unnatural in a \textit{classical} formula to insist that the
angular momentum be measured in units of Planck's constant. In our approach
(see (\ref{a.explicit}) below), angular momentum is measured in units of
$m\omega r^{2}.$ Although nothing prevents one from choosing $\omega$ so that
$m\omega r^{2}=\hbar,$ it seems artificial to us to insist on this. After all,
Eq. (6.1) concerns a classical construction that ought to be independent of
the value of Planck's constant.

We are now ready to apply the ``complexifier'' method of Thiemann \cite{T1}.
We take as our classical complexifier the kinetic energy function divided by
$\omega,$%
\begin{equation}
\text{complexifier}=\frac{\text{kinetic energy}}{\omega}=\frac{j^{2}}{2m\omega
r^{2}}.\label{complexifier.def}%
\end{equation}
We then define complex-valued functions $a_{1},\cdots,a_{d+1}$ on $T^{\ast
}(S^{d})$ by
\begin{align}
a_{k}\left(  \mathbf{x},\mathbf{p}\right)   & =\exp\left(  i\left\{
\cdot,\frac{j^{2}}{2m\omega r^{2}}\right\}  \right)  x_{k}\nonumber\\
& =\sum_{n=0}^{\infty}\left(  \frac{i}{2m\omega r^{2}}\right)  ^{n}\frac
{1}{n!}\underset{n}{\underbrace{\left\{  \cdots\left\{  \left\{  x_{k}%
,j^{2}\right\}  ,j^{2}\right\}  ,\cdots,j^{2}\right\}  }}.\label{a.def}%
\end{align}
Note that the $a_{k}$'s are obtained from the $x_{k}$'s by means of the
classical time-evolution generated by the kinetic energy function, evaluated
at the \textit{imaginary} time $i/\omega.$ The calculations below will show
that the series (\ref{a.def}) converges for all $\mathbf{x}$ and $\mathbf{p}. $

In \cite{T1} Thiemann allows \textit{any} function $C$ on the phase space to
be the complexifier, provided that $\exp(i\{\cdot,C\})x_{k}$ converges.
(Thiemann also allows any cotangent bundle to be the phase space.) The
condition of convergence, however, imposes severe restrictions on the choice
of $C,$ even when $C$ is quadratic in the momenta. We consider in this paper
only the complexifier (\ref{complexifier.def}).

To compute the functions $a_{k}$ explicitly, we first compute using (\ref{xj})
and (\ref{j2.def}) that, in vector notation,
\begin{equation}
\left\{  \mathbf{x},\frac{j^{2}}{2m\omega r^{2}}\right\}  =\frac{1}{m\omega
r^{2}}\mathbf{jx=}\frac{1}{m\omega}\mathbf{p}.\label{bracket}%
\end{equation}
On the other hand, it is easily verified that $\left\{  j_{kl},j^{2}\right\}
=0,$ which means that if we compute further Poisson brackets with $j^{2}$, the
matrix $\mathbf{j}$ gets ignored and we get
\[
\left(  \frac{1}{2m\omega r^{2}}\right)  ^{n}\underset{n}{\underbrace{\left\{
\cdots\left\{  \left\{  \mathbf{x},j^{2}\right\}  ,j^{2}\right\}
,\cdots,j^{2}\right\}  }}=\left(  \frac{1}{m\omega r^{2}}\right)
^{n}\mathbf{j}^{n}\mathbf{x.}%
\]
Here $\mathbf{j}^{n}\mathbf{x}$ means the matrix $\mathbf{j}$ applied $n$
times to the vector $\mathbf{x}.$

We obtain, then, the following ``polar coordinates'' expression for
$\mathbf{a}=\left(  a_{1},\cdots,a_{d+1}\right)  $%
\begin{equation}
\mathbf{a}\left(  \mathbf{x},\mathbf{p}\right)  =e^{i\mathbf{j}\left(
x,p\right)  /m\omega r^{2}}\mathbf{x}.\label{a.polar}%
\end{equation}
(Compare Eq. (3.37) in the first entry of \cite{T2}.) Using (\ref{jj2}) we can
compute this explicitly as
\begin{align}
\mathbf{a}\left(  \mathbf{x},\mathbf{p}\right)   & =\cosh\left(  \frac
{p}{m\omega r}\right)  \,\mathbf{x}+i\frac{r}{p}\sinh\left(  \frac{p}{m\omega
r}\right)  \mathbf{p}\nonumber\\
& =\cosh\left(  \frac{j}{m\omega r^{2}}\right)  \,\mathbf{x}+i\frac{r^{2}}%
{j}\sinh\left(  \frac{j}{m\omega r^{2}}\right)  \mathbf{p}.\label{a.explicit}%
\end{align}
We may at this point check the units: $p/m\omega r=j/m\omega r^{2}$ is
dimensionless and the whole expression has units of position. Note also that
$\mathbf{a}(\mathbf{x},-\mathbf{p})=\overline{\mathbf{a}\left(  \mathbf{x}%
,\mathbf{p}\right)  }.$

With $d=2$ (and $r=m\omega=1$) (\ref{a.explicit}) agrees with Eq. (6.1) of
\cite{KR1}. (See also Eq. (3.6) in the second entry of \cite{T2}.) In any
dimension (\ref{a.explicit}) agrees with the ``adapted complex structure'' on
$T^{\ast}(S^{d})$ as defined by Lempert and Sz\H{o}ke \cite{LS} and Guillemin
and Stenzel \cite{GStenz}, which for the special case of rank one symmetric
spaces was constructed earlier by Morimoto and Nagano \cite{MN}. See for
example p. 410 of \cite{Sz1}.

It is instructive to consider how this works out in the case of $S^{1}%
\subset\mathbb{R}^{2}.$ In that case we have only a single component of
angular momentum, $j_{12}=p_{1}x_{2}-p_{2}x_{1},$ so that $j=\left|
j_{12}\right|  .$ Since both terms in (\ref{a.explicit}) are even functions of
$j,$ we may replace $j$ by $j_{12}$ there. Then let $\theta$ be the usual
angular coordinate and let $\rho=-j_{12}/m\omega r^{2},$ so that $\rho$ is (up
to a constant) the canonically conjugate momentum to $\theta.$ Our phase space
is the set of points $\left(  x_{1},x_{2},p_{1},p_{2}\right)  $ with
$x_{1}^{2}+x_{2}^{2}=r^{2}$ and $p_{1}x_{1}+p_{2}x_{2}=0.$ On this set we have
the easily verified identity $j_{12}(x_{2},-x_{1})=r^{2}(p_{1},p_{2}).$ Upon
using this identity and $\mathbf{x}=r(\cos\theta,\sin\theta)$,
(\ref{a.explicit}) becomes
\begin{align}
\mathbf{a}  & =r(\cosh\rho\cos\theta-i\sinh\rho\sin\theta,\cosh\rho\sin
\theta+i\sinh\rho\cos\theta)\nonumber\\
& =r(\cos(\theta+i\rho),\sin(\theta+i\rho)).\label{a.s1}%
\end{align}
This result facilitates comparison with the analysis of the $S^{1}$ case in
\cite{KRP} and should be thought of as the ``complexification'' of the
identity $\mathbf{x}=r(\cos\theta,\sin\theta).$

As is well known, the Poisson bracket satisfies a Leibniz-type product rule,
$\left\{  f_{1},f_{2}f_{3}\right\}  =\left\{  f_{1},f_{2}\right\}  f_{3}%
+f_{2}\left\{  f_{1},f_{3}\right\}  ,$ and the analogous formula for Poisson
brackets, $\left\{  f_{1},\left\{  f_{2},f_{3}\right\}  \right\}  =\left\{
\left\{  f_{1},f_{2}\right\}  ,f_{3}\right\}  +\left\{  f_{2},\left\{
f_{1},f_{3}\right\}  \right\}  .$ (This last expression is equivalent to the
Jacobi identity.) Suppose then that we define the ``complexification''
$f_{\mathbb{C}}$ of any function $f$ to be
\[
f_{\mathbb{C}}=e^{i\left\{  \cdot,\text{ complexifier}\right\}  }f
\]
whenever the power series for the exponential converges. Then by a standard
power series argument we have
\begin{equation}
\left(  f_{1}f_{2}\right)  _{\mathbb{C}}=\left(  f_{1}\right)  _{\mathbb{C}%
}\left(  f_{2}\right)  _{\mathbb{C}}\label{complex.id1}%
\end{equation}
and
\begin{equation}
\left\{  f_{1},f_{2}\right\}  _{\mathbb{C}}=\left\{  \left(  f_{1}\right)
_{\mathbb{C}},\left(  f_{2}\right)  _{\mathbb{C}}\right\}
.\label{complex.id2}%
\end{equation}
Equation (\ref{complex.id1}) shows that if we ``complexify'' any polynomial in
the variables $x_{1},\cdots,x_{d+1}$ we will get simply the same polynomial in
$a_{1},\cdots,a_{d+1}.$ Furthermore, since $\left\{  x_{k},x_{l}\right\}  =0,$
(\ref{complex.id2}) shows that%

\begin{equation}
\left\{  a_{k},a_{l}\right\}  =0,\label{a.bracket}%
\end{equation}
which implies that $\left\{  \bar{a}_{k},\bar{a}_{l}\right\}  =0.$ The formula
for $\left\{  a_{k},\bar{a}_{l}\right\}  ,$ however, is complicated and we
will not compute it here.

Equation (\ref{complex.id1}) also shows that
\begin{equation}
a^{2}\left(  \mathbf{x},\mathbf{p}\right)  =r^{2}\label{a.squared}%
\end{equation}
for all $\mathbf{x},\mathbf{p}$, which is also evident from (\ref{a.explicit}%
). Thus the map $\left(  \mathbf{x},\mathbf{p}\right)  \rightarrow
\mathbf{a}\left(  \mathbf{x},\mathbf{p}\right)  $ defines a map of the
cotangent bundle $T^{\ast}(S^{d})$ to the \textit{complexified sphere}
\begin{equation}
S_{\mathbb{C}}^{d}=\left\{  \left.  \mathbf{a}\in\mathbb{C}^{d+1}\right|
a_{1}^{2}+\cdots+a_{d+1}^{2}=r^{2}\right\}  .\label{sdc.def}%
\end{equation}
It is not hard to see that this map is invertible, indeed a diffeomorphism of
$T^{\ast}(S^{d})$ with $S_{\mathbb{C}}^{d}.$

\section{The annihilation operators\label{annihilate.sec}}

We now consider the quantum counterpart of the constructions in the previous
section. This means that the functions $j_{kl}$ and $x_{k}$ should be replaced
by self-adjoint operators $J_{kl}$ and $X_{k}$ acting on (suitable domains in)
some separable complex Hilbert space. These should satisfy $J_{lk}=-J_{kl}$
and the quantum counterpart of the Poisson-bracket relations (\ref{xj}),
namely,
\begin{align}
\frac{1}{i\hbar}\left[  J_{kl},J_{mn}\right]   & =\delta_{kn}J_{lm}%
+\delta_{lm}J_{kn}-\delta_{km}J_{l{}n}-\delta_{l{}n}J_{km}\nonumber\\
\frac{1}{i\hbar}\left[  X_{k},J_{lm}\right]   & =\delta_{kl}X_{m}-\delta
_{km}X_{l}\label{qxj}\\
\frac{1}{i\hbar}\left[  X_{k},X_{l}\right]   & =0\nonumber
\end{align}
We recognize this as a representation of the Euclidean Lie algebra
$\mathrm{e}(d+1)=\mathrm{so}(d+1)\ltimes\mathbb{R}^{d+1}.$ We assume that this
representation of $\mathrm{e}(d+1)$ comes from an irreducible unitary
representation of the associated connected, simply connected Lie group
$\mathrm{\tilde{E}}(d+1).$ Here $\mathrm{\tilde{E}}(d+1)\cong\mathrm{Spin}%
(d+1)\ltimes\mathbb{R}^{d+1}$ for $d\geq2$ and $\mathrm{\tilde{E}}%
(2)\cong\mathbb{R}\ltimes\mathbb{R}^{2},$ where $\ltimes$ denotes a semidirect
product with the normal factor on the right.

The irreducible unitary representations of $\mathrm{\tilde{E}}(d+1)$ may be
classified by the Wigner--Mackey method. One first chooses an orbit of
$\mathrm{Spin}(d+1)$ on $\mathbb{R}^{d+1},$ namely, a sphere of some radius
$r\geq0.$ Since the case $r=0$ is presumably unphysical (though mathematically
permitted), we assume from now on that $r>0.$ Next one selects any one point
in the sphere of radius $r$ and considers the ``little group,'' that is, the
stabilizer in $\mathrm{Spin}(d+1)$ of the point. For $r>0$ the little group is
simply $\mathrm{Spin}(d).$ The irreducible representations of $\mathrm{\tilde
{E}}(d+1)$ are then labeled by the value of $r$ and by an irreducible unitary
representation of the little group. In this paper we will consider only the
case in which the representation of the little group is trivial. Nevertheless
the definitions of the annihilation operators and of the coherent states make
sense in general.

Choosing a sphere of radius $r$ amounts to requiring that the operators in
(\ref{qxj}) satisfy
\begin{equation}
X^{2}=r^{2}.\label{qconstraint.1}%
\end{equation}
We will shortly impose an additional condition among the $X$'s and $J$'s that
forces the representation of the little group to be trivial. For now, however,
we will assume only the $\mathrm{e}(d+1)$ relations (\ref{qxj}) and the
condition (\ref{qconstraint.1}).

We define the total angular momentum $J^{2}$ as in the classical case by
\begin{equation}
J^{2}=\sum_{k<l}J_{kl}^{2}.\label{qj.squared}%
\end{equation}
As in the classical case we define
\[
\text{complexifier}=\frac{\text{kinetic energy}}{\omega}=\frac{J^{2}}{2m\omega
r^{2}}.
\]

We then define the annihilation operators by replacing $\left\{
\cdot,\text{complexifier}\right\}  $ in (\ref{a.def}) with its quantum
counterpart, $(1/i\hbar)\left[  \cdot,\text{complexifier}\right]  $:
\begin{align}
A_{k}  & =\exp\left(  \frac{i}{i\hbar}\left[  \cdot,\frac{J^{2}}{2m\omega
r^{2}}\right]  \right)  X_{k}\nonumber\\
& =\sum_{n=0}^{\infty}\frac{1}{(2m\omega r^{2}\hbar)^{n}}\frac{1}{n!}\left[
\cdots\left[  \left[  X_{k},J^{2}\right]  ,J^{2}\right]  \cdots,J^{2}\right]
.\label{qa.def}%
\end{align}
By a standard formula from Lie group theory this may be written as
\begin{equation}
A_{k}=e^{-J^{2}/2m\omega r^{2}\hbar}X_{k}e^{J^{2}/2m\omega r^{2}\hbar
}.\label{qa.commutator}%
\end{equation}
In the general form of Thiemann's method, (\ref{qa.commutator}) would be 
$\exp(-\hat{C}/\hbar)X_{k}\exp(\hat{C}/\hbar),$ where $\hat{C}$ is the
quantum operator corresponding to the classical complexifier $C.$

For determining the eigenvectors of the annihilation operators (i.e. the
coherent states), (\ref{qa.commutator}) is the most useful expression for the
$A_{k}$'s. Nevertheless we will give two other formulas, a polar decomposition
and an ``explicit'' formula in terms of the position and momentum operators.
The $A_{k}$'s are unbounded operators and so something must be said about
their domains; see the discussion at the end of this section. The annihilation
operators satisfy (in analogy to (\ref{a.bracket}) and (\ref{a.squared}))
\begin{align}
\left[  A_{k},A_{l}\right]   & =0\label{qa.bracket}\\
A^{2}  & =r^{2}\label{qa.squared}%
\end{align}
(since $\left[  X_{k},X_{l}\right]  =0$ and $X^{2}=r^{2}$).

To compute $\mathbf{A}$ we first compute using (\ref{qxj}) and
(\ref{qj.squared}) that
\begin{align*}
\frac{1}{i\hbar}\left[  X_{k},J^{2}\right]   & =J_{kl}X_{l}+X_{l}J_{kl}\\
& =2J_{kl}X_{l}+i\hbar\left(  \delta_{lk}X_{l}-\delta_{ll}X_{k}\right) \\
& =2J_{kl}X_{l}-i\hbar dX_{k}.
\end{align*}
Here we have chosen to order things with the $J$'s to the left of the $X$'s
and we use the sum convention. Thus in vector notation we have
\begin{equation}
\frac{1}{i\hbar}\left[  \mathbf{X},\frac{J^{2}}{2m\omega r^{2}}\right]
=\frac{1}{m\omega r^{2}}\left(  \mathbf{J}-\frac{i\hbar d}{2}\right)
\mathbf{X}.\label{qbracket}%
\end{equation}
Here the term involving $i\hbar d/2$ is a ``quantum correction''; compare
(\ref{bracket}).

Since $\left[  J_{kl},J^{2}\right]  =0,$ further brackets will give just
another factor of the matrix operator $\mathbf{J}-i\hbar d/2.$ Thus the polar
coordinates decomposition of $\mathbf{A}$ has just a single quantum
correction, namely,
\begin{equation}
\mathbf{A}=\exp\left\{  \frac{i\mathbf{J}+\hbar d/2}{m\omega r^{2}}\right\}
\mathbf{X}\label{qpolar}%
\end{equation}
In the case $d=2,$ a formula very similar to this is taken in \cite{KR1} as
the definition of the annihilation operators. The only difference is that
Kowalski and Rembieli\'{n}ski formulate things in terms of $2\times2$ matrix
operators, whereas in the case $d=2$ (\ref{qpolar}) is in terms of $3\times3$
matrix operators. Nevertheless, our expression is equivalent to that of
Kowalski and Rembieli\'{n}ski; see below and Eq. (5.9) of \cite{KR2}. The
analog of (\ref{qpolar}) for the group case is given in Eq. (3.44) of the
first entry in \cite{T2} and in Eq. (3.13) of the second entry in \cite{T2}.

Note that the definition (\ref{qa.def}) makes sense in any irreducible
representation of $\mathrm{\tilde{E}}(d+1)$ (with $X^{2}=r^{2}>0$), and that
the formula (\ref{qpolar}) is valid in this generality. However, to compute
$\mathbf{A}$ more explicitly than this we need to further specify the
irreducible representation of $\mathrm{\tilde{E}}(d+1).$ We limit ourselves to
the case in which the representation of the little group $\mathrm{Spin}(d) $
is trivial. This corresponds to a quantum particle on the sphere with no
internal degrees of freedom. In the case the case of $S^{2},$ this corresponds
to taking the ``twist'' (in the notation of Kowalski and Rembieli\'{n}ski) to
be zero. We show in Section \ref{euclid.sec} that the little group acts
trivially if and only if the following relation holds:
\begin{equation}
X^{2}J_{kl}=J_{km}X_{m}X_{l}-J_{lm}X_{m}X_{k}.\label{qconstraint.2}%
\end{equation}
This is the quantum counterpart of the classical constraint
(\ref{constraint.2}).

In computing $\mathbf{A}$ it is convenient to introduce ``momentum'' operators
$P_{k}$ given by
\[
P_{k}:=\frac{J_{kl}X_{l}}{r^{2}}.
\]
These operators are \textit{not} self-adjoint and we have chosen to put the
$J$'s to the left of the $X$'s (because we have put $\mathbf{J}$ to the left
of $\mathbf{X}$ in (\ref{qbracket}) and (\ref{qpolar})). We may re-write
(\ref{qconstraint.2}) in terms of the $P_{k}$'s as
\begin{equation}
J_{kl}=P_{k}X_{l}-P_{l}X_{k}.\label{qconstraint.3}%
\end{equation}
The position and momentum operators satisfy
\begin{equation}
\frac{1}{i\hbar}\left[  X_{k},P_{l}\right]  =\delta_{kl}I-\frac{X_{k}X_{l}%
}{r^{2}}.\label{qxp.bracket}%
\end{equation}
(Compare (\ref{xp.bracket}).) We may also compute using (\ref{qxj}) the
quantum counterpart of $\mathbf{x}\cdot\mathbf{p}=0,$ which is really two
relations on the quantum side:
\begin{align*}
\mathbf{P}\cdot\mathbf{X}  & =0\\
\mathbf{X}\cdot\mathbf{P}  & =i\hbar dI.
\end{align*}

We now write down the formulas that allow us to compute $\mathbf{A}$ in terms
of $\mathbf{X}$ and $\mathbf{P}$:
\begin{align}
\mathbf{JX}  & =r^{2}\mathbf{P}\nonumber\\
\mathbf{JP}  & =-P^{2}\mathbf{X}+i\hbar(d-1)\mathbf{P}.\label{jj22}%
\end{align}
The first line is simply the definition of $\mathbf{P}$. The second line comes
from (\ref{qconstraint.2}) or (\ref{qconstraint.3}) and is essential to the
explicit calculation of the annihilation operators in terms of $\mathbf{X}$
and $\mathbf{P}.$ Note that there is an additional quantum correction here. To
verify the second line of (\ref{jj22}), write $\mathbf{J} $ in terms of
$\mathbf{P}$ using (\ref{qconstraint.3}) and then use (\ref{qxp.bracket}).

We now treat $\mathbf{J}$ as a $2\times2$ matrix acting on the ``basis''
$\mathbf{X}$ and $\mathbf{P},$ as given in (\ref{jj22}). Since all the entries
of this $2\times2$ matrix commute, we can just treat $P^{2}$ as a scalar and
compute an ordinary $2\times2$ matrix exponential. So effectively we have
\[
\mathbf{J}=\left(
\begin{array}
[c]{cc}%
0 & -P^{2}\\
r^{2} & i\hbar(d-1)
\end{array}
\right)  .
\]
One can then compute the exponential of this matrix either by hand or using a
computer algebra program. A calculation shows that $P^{2}=r^{-2}J^{2}$ as in
the classical case. It is convenient to express things in terms of the scalar
operator
\[
J:=\sqrt{J^{2}+\hbar^{2}(d-1)^{2}/4}.
\]
Then after exponentiating $\mathbf{J,}$ (\ref{qpolar}) becomes
\begin{align}
\mathbf{A}  & =e^{\hbar/2m\omega r^{2}}\cosh\left(  \frac{J}{m\omega r^{2}%
}\right)  \mathbf{X}+e^{\hbar/2m\omega r^{2}}\frac{\hbar(d-1)}{2J}\sinh\left(
\frac{J}{m\omega r^{2}}\right)  \mathbf{X}\nonumber\\
& +ie^{\hbar/2m\omega r^{2}}\frac{r^{2}}{J}\sinh\left(  \frac{J}{m\omega
r^{2}}\right)  \mathbf{P}.\label{qa.explicit}%
\end{align}

Equation (\ref{qa.explicit}) is similar to the corresponding classical
expression (\ref{a.explicit}), with only the following differences: 1) there
is an overall factor of $\exp(\hbar/2m\omega r^{2}),$ 2) the quantity $j$ in
(\ref{a.explicit}) is replaced by $(J^{2}+\hbar^{2}(d-1)^{2}/4)^{1/2},$ and 3)
there is an extra $\sinh$ term in the coefficient of $\mathbf{X}$ that does
not occur in the classical formula. Note that the above expression formally
coincides with the classical one in the limit $\hbar\rightarrow0.$ In the case
$d=2$ with $r=m\omega=\hbar=1$ (\ref{qa.explicit}) agrees with Eq. (4.16) in
\cite{KR1}. In the case $d=3$ (identifying $S^{3}$ with $\mathrm{SU}(2)$ and
adjusting for minor differences of normalization) (\ref{qa.explicit}) agrees
with Eq. (3.132) in the last entry in \cite{T2}. In the case $d=1$ we get an
expression identical to the classical expression (\ref{a.s1}) except for an
overall factor of $\exp(\hbar/m\omega r^{2})$ (compare Eqs. (3.3) and (3.4) of
\cite{KRP}).

It is clear from (\ref{qa.explicit}) that the $A_{k}$'s are unbounded
operators, as expected since the $a_{k}$'s are unbounded functions. This means
that the $A_{k}$'s cannot be defined on the whole Hilbert space, but only on
some dense subspace, which should be specified. We take the expression
(\ref{qa.commutator}) as our definition of the annihilation operators. We
first define the $A_{k}$'s on what we will call the ``minimal domain,''
namely, the space of finite linear combinations of spherical harmonics (that
is, of eigenvectors for $J^{2}$). The expression (\ref{qa.commutator}) makes
sense on the minimal domain, since each of the three factors making up $A_{k}$
preserve this space. We consider also a ``maximal domain'' for the $A_{k}$'s,
defined as follows. Given any vector $\left|  \phi\right\rangle $ in the
Hilbert space, we expand $\left|  \phi\right\rangle $ in a series expansion in
terms of spherical harmonics. Then we apply $A_{k}$ term-by-term, that is, by
formally interchanging $A_{k}$ with the sum. The result will then be a formal
series of spherical harmonics. If this formal series converges in the Hilbert
space then we say that $\left|  \phi\right\rangle $ is in the maximal domain
of $A_{k}$ and that the value of $A_{k}\left|  \phi\right\rangle $ is the sum
of this series. (It can be shown that the product of $x_{k}$ and a spherical
harmonic of degree $n$ is the sum of a spherical harmonic of degree $n+1$ and
a spherical harmonic of degree $n-1.$ It follows that the degree $l$ term in
the expansion of $A_{k}\left|  \phi\right\rangle $ involves only the degree
$n-1$ and degree $n+1$ terms of $\left|  \phi\right\rangle .$ So each term in
the formal series for $A_{k}\left|  \phi\right\rangle $ can be computed by
means of a finite sum.)

It can be shown that if one starts with the operator $A_{k}$ on its minimal
domain and then takes its closure (in the functional analytic sense) the
result is the operator $A_{k}$ on its maximal domain. Thus if we want $A_{k}$
to be a closed operator there is only one reasonable choice for its domain.
The coherent states will not be finite linear combinations of spherical
harmonics but will be in the maximal domain of all the $A_{k}$'s.

\section{The coherent states\label{coherent.sec}}

We are now ready to introduce the coherent states, which we define to be the
simultaneous eigenvectors of the annihilation operators. These coherent states
are \textit{not of Perelomov type}. Although we have described the quantum
Hilbert space as an irreducible representation of $\mathrm{\tilde{E}}(d+1),$
the coherent states are not obtained from one fixed vector by the action of
$\mathrm{\tilde{E}}(d+1).$ Indeed the only elements of $\mathrm{\tilde{E}%
}(d+1)$ that preserve the set of coherent states are the rotations. See
Section \ref{conclusions.sec} for a comparison of these coherent states to the
generalized Perelomov-type coherent states for $\mathrm{\tilde{E}}(d+1),$ as
constructed either by De Bi\`{e}vre or by Isham and Klauder.

The coherent states will be simultaneous eigenvectors of the annihilation
operators $A_{k}$, and thus can be thought of as the quantum counterparts of a
classical state with definite values for the complex coordinates $a_{k}.$ On
the quantum side, however, $A_{k}^{\dagger}$ does not commute with $A_{k},$
and thus although the coherent states satisfy $A_{k}\left|  \psi\right\rangle
=a_{k}\left|  \psi\right\rangle $ they do \textit{not} satisfy $A_{k}%
^{\dagger}\left|  \psi\right\rangle =\bar{a}_{k}\left|  \psi\right\rangle .$

We use the formula (\ref{qa.commutator}) for the annihilation operators. If we
introduce the dimensionless form of the total angular momentum,
\[
\tilde{J}^{2}=\frac{1}{\hbar^{2}}J^{2},
\]
then this may be expressed as
\begin{equation}
\mathbf{A}=e^{-\tau\tilde{J}^{2}/2}\mathbf{X}e^{\tau\tilde{J}^{2}%
/2},\label{qa.dimensionless}%
\end{equation}
where $\tau$ is the dimensionless quantity given by
\[
\tau=\frac{\hbar}{m\omega r^{2}}.
\]

The parameter $\tau$ is a new feature of the sphere case; no such
dimensionless quantity arises in the $\mathbb{R}^{d}$ case. The significance
of $\tau$ for the coherent states is that it controls the ratio of the spatial
width of the coherent states to the radius of the sphere. Specifically, we
expect the approximate spatial width $\Delta X$ of a coherent state to be
$\sqrt{\hbar/2m\omega},$ at least if this quantity is small compared to $r.$
In that case
\[
\frac{\Delta X}{r}\approx\frac{\sqrt{\hbar/2m\omega}}{r}=\sqrt{\frac{\tau}{2}%
}.
\]
So if $\tau\ll1$ we expect the coherent states to be concentrated in a small
portion of the sphere, and to look, in appropriate coordinates, approximately
Gaussian. This has been proved \cite{T2} for the case of $S^{3}=\mathrm{SU}(2).$

Kowalski and Rembieli\'{n}ski implicitly take $\tau=1$ in their treatment of
the $d=2$ case, since they choose units with $m=r=\hbar=1,$ and since they do
not have the parameter $\omega.$ (See our comments in Section
\ref{complex.sec} about the parameter $\omega.$) To us this seems a needless
loss of generality, even though it is easy to insert $\tau$ in the appropriate
places in their formulas.

We now proceed with the construction of the coherent states. For each
$\mathbf{a}$ in the \textit{real} sphere $S^{d},$ let $\left|  \delta
_{\mathbf{a}}\right\rangle $ be the (generalized) position eigenfunction with
$X_{k}\left|  \delta_{\mathbf{a}}\right\rangle =a_{k}\left|  \delta
_{\mathbf{a}}\right\rangle .$ Since we assume that the little group acts
trivially these position eigenfunctions are (for each $\mathbf{a}$) unique up
to a constant and we may normalize them so that the action of the rotation
group takes $\left|  \delta_{\mathbf{a}}\right\rangle $ to $\left|
\delta_{R\mathbf{a}}\right\rangle ,$ $R\in\mathrm{SO}(d+1).$ If we let
\begin{equation}
\left|  \psi_{\mathbf{a}}\right\rangle =e^{-\tau\tilde{J}^{2}/2}\left|
\delta_{\mathbf{a}}\right\rangle .\label{first.states}%
\end{equation}
then it follows immediately from (\ref{qa.dimensionless}) that $\left|
\psi_{\mathbf{a}}\right\rangle $ is a simultaneous eigenvector for each
$A_{k}$ with eigenvalue $a_{k}.$ Although $\left|  \delta_{\mathbf{a}%
}\right\rangle $ is non-normalizable, the smoothing nature of the operator
$\exp(-\tau\tilde{J}^{2}/2)$ guarantees that $\left|  \psi_{\mathbf{a}%
}\right\rangle $ is normalizable for all $\mathbf{a}\in S^{d}.$ A key result
of this section is the following proposition, which asserts that we can
analytically continue the coherent states $\left|  \psi_{\mathbf{a}%
}\right\rangle $ with respect to $\mathbf{a}$ so as to obtain states labeled
by points $\mathbf{a}$ in the \textit{complex} sphere $S_{\mathbb{C}}^{d}.$

\begin{proposition}
\label{continue.prop}There exists a unique family of states $\left|
\psi_{\mathbf{a}}\right\rangle $ parameterized by $\mathbf{a}\in
S_{\mathbb{C}}^{d} $ such that 1) the states depend holomorphically on
$\mathbf{a},$ and 2) for $\mathbf{a}\in S^{d},$ they agree with the states in
(\ref{first.states}). These are normalizable states and satisfy
\[
A_{k}\left|  \psi_{\mathbf{a}}\right\rangle =a_{k}\left|  \psi_{\mathbf{a}%
}\right\rangle ,\quad\mathbf{a}\in S_{\mathbb{C}}^{d}.
\]
\end{proposition}

We call these states the \textit{coherent states}. Note that we have then one
coherent state for each point in $S_{\mathbb{C}}^{d},$ that is, one coherent
state for each point in the classical phase space. It can be shown that these
are (up to a constant) the \textit{only} simultaneous eigenvectors of the
annihilation operators. These coherent states are \textit{not} normalized to
be unit vectors. The proof of Proposition \ref{continue.prop} is at the end of
this section.

Note that since the operator $\tilde{J}^{2}$ commutes with rotations, the
action of the rotation subgroup $\mathrm{SO}(d+1)$ of $\mathrm{E}(d+1)$ will
take $\left|  \psi_{\mathbf{a}}\right\rangle $ to $\left|  \psi_{R\mathbf{a}%
}\right\rangle $ for any $R\in\mathrm{SO}(d+1).$ On sufficiently regular
states we can analytically continue the action of $\mathrm{SO}(d+1)$ to an
action of $\mathrm{SO}(d+1;\mathbb{C}),$ which will take $\left|
\psi_{\mathbf{a}}\right\rangle $ to $\left|  \psi_{R\mathbf{a}}\right\rangle $
for any $R\in\mathrm{SO}(d+1;\mathbb{C}).$ Then any coherent state can be
obtained from any other by the action of $\mathrm{SO}(d+1;\mathbb{C}).$ Since,
however, the action of $\mathrm{SO}(d+1;\mathbb{C})$ is neither unitary nor
irreducible, this observation still does not bring the coherent states into
the Perelomov framework.

We can give an explicit formula for the coherent states in the position
representation in terms of the \textit{heat kernel} on $S^{d}.$ The heat
kernel is the function on $S^{d}\times S^{d}$ given by $\rho_{\tau}%
(\mathbf{x},\mathbf{y})=\left\langle \delta_{\mathbf{x}}\right|
e^{-\tau\tilde{J}^{2}/2}\left|  \delta_{\mathbf{y}}\right\rangle .$ It can be
shown (see \cite{H1} or the formulas below) the $\rho_{\tau}$ extends
(uniquely) to a holomorphic function on $S_{\mathbb{C}}^{d}\times
S_{\mathbb{C}}^{d},$ also denoted $\rho_{\tau}.$ In terms of the analytically
continued heat kernel the coherent states are given by
\begin{equation}
\left\langle \delta_{\mathbf{x}}|\psi_{\mathbf{a}}\right\rangle =\rho_{\tau
}(\mathbf{a},\mathbf{x}),\quad\mathbf{a}\in S_{\mathbb{C}}^{d},\,\mathbf{x}\in
S^{d}.\label{position.formula}%
\end{equation}

Meanwhile, explicit formulas for the heat kernel may be found, for example, in
\cite{Ta,Ca}. For $\mathbf{x}$ and $\mathbf{y}$ in the real sphere,
$\rho_{\tau}(\mathbf{x},\mathbf{y})$ depends only on the angle $\theta$
between $\mathbf{x}$ and $\mathbf{y},$ where $\theta=\cos^{-1}(\mathbf{x}%
\cdot\mathbf{y}/r^{2}).$ This remains true for $\rho_{\tau}(\mathbf{a}%
,\mathbf{x}),$ with $\mathbf{a}\in S_{\mathbb{C}}^{d}$, except now
$\theta=\cos^{-1}(\mathbf{a}\cdot\mathbf{x}/r^{2})$ is complex-valued. Of
course the inverse cosine function is multiple-valued, but because the heat
kernel is an even, $2\pi$-periodic function of $\theta,$ it does not matter
which value of $\theta$ we use, provided that $\cos\theta=\mathbf{a}%
\cdot\mathbf{x}/r^{2}.$

We now record the formulas, writing $\rho_{\tau}^{d}$ to indicate the
dependence on the dimension. For $d=1,2,3$ we have
\begin{align*}
\rho_{\tau}^{1}(\mathbf{a},\mathbf{x})  & =(2\pi\tau)^{-1/2}\sum_{n=-\infty
}^{\infty}e^{-(\theta-2\pi n)^{2}/2\tau}\\
\rho_{\tau}^{2}(\mathbf{a},\mathbf{x})  & =(2\pi\tau)^{-1}e^{\tau/8}\frac
{1}{\sqrt{\pi\tau}}\int_{\theta}^{\pi}\frac{1}{\sqrt{\cos\theta-\cos\phi}}%
\sum_{n=-\infty}^{\infty}(-1)^{n}(\phi-2\pi n)e^{-(\phi-2\pi n)^{2}/2\tau
}\,d\phi\\
\rho_{\tau}^{3}(\mathbf{a},\mathbf{x})  & =(2\pi\tau)^{-3/2}e^{\tau/2}\frac
{1}{\sin\theta}\sum_{n=-\infty}^{\infty}(\theta-2\pi n)e^{-(\theta-2\pi
n)^{2}/2\tau}.
\end{align*}
In the formula for $\rho_{\tau}^{2}$ we may without loss of generality take
$\theta$ with $0\leq\operatorname{Re}\theta\leq\pi,$ in which case the
integral is to be interpreted as a contour integral in the strip
$0\leq\operatorname{Re}\phi\leq\pi.$ The relatively simple formula for the
heat kernel on $S^{3}=\mathrm{SU}(2)$ allows for detailed calculations for the
coherent states in this case, as carried out in \cite{T2}. To find the formula
in higher dimensions we use the inductive formula
\[
\rho_{\tau}^{d+2}(\mathbf{a},\mathbf{x})=-e^{d\tau/2}\frac{1}{2\pi\sin\theta
}\frac{d}{d\theta}\rho_{\tau}^{d}(\mathbf{a},\mathbf{x}).
\]

There is also an expression for the heat kernel in terms of spherical
harmonics. For example, when $d=2$ we have
\begin{equation}
\rho_{\tau}^{2}(\mathbf{a},\mathbf{x})=\sum_{l=0}^{\infty}e^{-\tau
l(l+1)/2}\sqrt{2l+1}P_{l}(\cos\theta),\label{coherent.expand}%
\end{equation}
where the $P_{l}$ is the Legendre polynomial of degree $l.$ (Compare Eq. (5.3)
of \cite{KR1}.) The earlier expression for $\rho_{\tau}^{2}$ is a sort of
Poisson-summed version of (\ref{coherent.expand})--see \cite{Ca}.

We also consider the reproducing kernel, defined by
\begin{equation}
R_{\tau}(\mathbf{a,b})=\left\langle \left.  \psi_{\mathbf{b}}\right|
\psi_{\mathbf{a}}\right\rangle ,\quad\mathbf{a},\mathbf{b}\in S_{\mathbb{C}%
}^{d}.\label{repro.def}%
\end{equation}
In terms of the analytically continued heat kernel the reproducing kernel is
given by
\[
R_{\tau}(\mathbf{a,b})=\rho_{2\tau}(\mathbf{a},\mathbf{\bar{b}}),\quad
\mathbf{a},\mathbf{b}\in S_{\mathbb{C}}^{d}.
\]
Note that $R_{\tau}(\mathbf{a},\mathbf{b})$ depends holomorphically on
$\mathbf{a}$ and anti-holomorphically on $\mathbf{b}.$

In the case of $S^{1}=\mathrm{U}(1)$ and $S^{3}=\mathrm{SU}(2),$ Thiemann and
Winkler have proved in the second and third entries of \cite{T2} that the
coherent states defined here satisfy good phase space localization properties
and that the Ehrenfest theorem holds infinitesimally. We fully expect that
these results hold for all $d.$ This expectation is based on the idea that the
heat kernel in (\ref{position.formula}) will behave for small $\tau$ like
$c\exp(-\theta^{2}/2\tau),$ even for complex values of $\theta.$ Thiemann and
Winkler have verified this in the cases $d=1,3$ and a similar analysis should
be possible in general, using the explicit formulas for small $d$ and the
inductive formula for $\rho^{d+2}$ in terms of $\rho^{d}.$

\textit{Proof of Proposition \ref{continue.prop}}. There are two ways to prove
this proposition. The simplest way is to use the expression for $\psi
_{\mathbf{a}}$ in terms of the heat kernel $\rho_{\tau}$ and the explicit
formulas above for $\rho_{\tau}.$ It is easily seen that $\rho_{\tau}$ extends
to an entire holomorphic function of $\theta.$ Thus the expression
$\left\langle \delta_{\mathbf{x}}|\psi_{\mathbf{a}}\right\rangle =\rho_{\tau
}(\mathbf{a},\mathbf{x})$ makes sense for any $\mathbf{a}$ in $S_{\mathbb{C}%
}^{d},$ with $\cos\theta$ and thus also $\theta$ taking complex values. It is
not hard to see that the $\left|  \psi_{\mathbf{a}}\right\rangle $, so
defined, is in the (maximal) domain of the annihilation operators and that it
depends holomorphically on $\mathbf{a}\in S_{\mathbb{C}}^{d}.$ Since
$A_{k}\left|  \psi_{\mathbf{a}}\right\rangle =a_{k}\left|  \psi_{\mathbf{a}%
}\right\rangle $ for $\mathbf{a}\in S^{d},$ an analytic continuation argument
will show that this equation remains true for all $\mathbf{a}\in
S_{\mathbb{C}}^{d}.$ Alternatively we may use the expansion of the coherent
states in terms of spherical harmonics as in (\ref{coherent.expand}) and show
that this expression can be analytically continued term-by-term in
$\mathbf{a}$. (Compare Section 4 of \cite{H1}.) $\square$

\section{The resolution of the identity\label{res.sec}}

We now choose a coordinate system in which $r=1$ and $m\omega=1$. This amounts
to using the normalized position $\mathbf{x}/r$ and normalized momentum
$\mathbf{p}/m\omega r.$ Since these choices set our position and momentum
scales we cannot also take $\hbar=1.$ Note that the dimensionless parameter
$\tau=\hbar/m\omega r^{2}$ equals $\hbar$ in such a coordinate system. We now
write $\left|  \psi_{\mathbf{x},\mathbf{p}}\right\rangle $ for $\left|
\psi_{\mathbf{a}(\mathbf{x},\mathbf{p})}\right\rangle $.

\begin{theorem}
\label{res.thm}The coherent states have a resolution of the identity of the
form
\begin{equation}
I=\int_{\mathbf{x}\in S^{d}}\int_{\mathbf{p}\cdot\mathbf{x}=0}\left|
\psi_{\mathbf{x},\mathbf{p}}\right\rangle \left\langle \psi_{\mathbf{x}%
,\mathbf{p}}\right|  \,\nu\left(  2\tau,2p\right)  \left(  \frac{\sinh2p}%
{2p}\right)  ^{d-1}\,2^{d}d\mathbf{p}\,d\mathbf{x}\label{res.int}%
\end{equation}
where $\nu(s,R)$ is the solution to the differential equation
\[
\frac{d\nu(s,R)}{ds}=\frac{1}{2}\left[  \frac{\partial^{2}\nu}{\partial R^{2}%
}-(d-1)\frac{\cosh R}{\sinh R}\frac{\partial\nu}{\partial R}\right]
\]
subject to the initial condition
\[
\lim_{s\downarrow0}\,c_{d}\int_{0}^{\infty}f(R)\nu(s,R)(\sinh R)^{d-1}%
\,dR=f(0)
\]
for all continuous functions $f$ on $\left[  0,\infty\right)  $ with at most
exponential growth at infinity. Here $d\mathbf{x}$ is the surface area measure
on $S^{d},$ $\tau$ is the dimensionless quantity $\tau=\hbar/m\omega r^{2},$
and $c_{d}$ is the volume of the unit sphere in $\mathbb{R}^{d}.$
\end{theorem}

The operator on the right side of the equation for $\nu$ is just the radial
part of the Laplacian for $d$-dimensional hyperbolic space \cite[Sect.
5.7]{Da}. This means that $\nu(s,R)$ is the heat kernel for hyperbolic space,
that is, the fundamental solution of the heat equation. Hyperbolic space is
the non-compact, negatively curved ``dual'' of the compact, positively curved
symmetric space $S^{d}.$ Note that the function $\nu$ is evaluated at ``time''
$2\tau$ and radius $2p.$ The inversion formula for the Segal--Bargmann
transform, described in Section \ref{sb.sec}, involves the function $\nu$
evaluated at time $\tau$ and radius $p.$

The resolution of the identity for the coherent states will be obtained by
continuously varying the dimensionless parameter $\tau.$ When $\tau=0$ the
coherent states are simply the position eigenvectors, which have a resolution
of the identity because the position operators are self-adjoint. We will show
that the function $\nu$ satisfies the correct differential equation to make
the resolution of the identity remain true as we move to non-zero $\tau.$

Theorem \ref{res.thm} is a special case of Theorem 3 of \cite{St}, written out
more explicitly and re-stated in terms of coherent states instead of the
Segal--Bargmann transform. However we give below a self-contained and
elementary proof. The case $d=2$ is also described (with a different proof) in
\cite{KR2}. Since $S^{3}=\mathrm{SU}(2),$ the $d=3$ case belongs to the group
case, which is found in \cite{H1}. See also Section 4.4 of the second entry in
\cite{T2} for another proof in the $\mathrm{SU}(2)$ case.

We report here the formulas for the function $\nu(s,R),$ which may be found,
for example, in \cite[Sect. 5.7]{Da} or \cite[Eq. (8.73)]{Ca}. Writing
$\nu_{d}(s,R)$ to make explicit the dependence on the dimension we have
\begin{align*}
\nu_{1}(s,R)  & =(2\pi s)^{-1/2}\,e^{-R^{2}/2s}\\
\nu_{2}(s,R)  & =(2\pi s)^{-1}e^{-s/8}\frac{1}{\sqrt{\pi s}}\int_{R}^{\infty
}\frac{\rho e^{-\rho^{2}/2s}}{\left(  \cosh\rho-\cosh R\right)  ^{1/2}}%
\,d\rho\\
\nu_{3}(s,R)  & =(2\pi s)^{-3/2}e^{-s/2}\frac{R}{\sinh R}e^{-R^{2}/2s}.
\end{align*}
and the recursion relation
\[
\nu_{d+2}(s,R)=-\frac{e^{-ds/2}}{2\pi\sinh R}\frac{\partial}{\partial R}%
\nu_{d}(s,R).
\]
Estimates on the behavior as $R\rightarrow\infty$ of $\nu$ may be found in
\cite[Sect 5.7]{Da} and in \cite{DM}. Note the similarities between the
formulas for $\nu$ and the formulas for the heat kernel $\rho_{\tau}$ on the sphere.

Some care must be taken in the interpretation of the integral (\ref{res.int}).
Even in the $\mathbb{R}^{d}$ case this integral is not absolutely convergent
in the operator norm sense. Rather the appropriate sense of convergence is the
weak sense. This means that for all vectors $\phi_{1},\phi_{2}$ in the Hilbert
space we have
\begin{equation}
\left\langle \phi_{1}|\phi_{2}\right\rangle =\int_{\mathbf{x}\in S^{d}}%
\int_{\mathbf{p}\cdot\mathbf{x}=0}\left\langle \phi_{1}|\psi_{\mathbf{x}%
,\mathbf{p}}\right\rangle \left\langle \psi_{\mathbf{x},\mathbf{p}}|\phi
_{2}\right\rangle \,\nu\left(  2\tau,2p\right)  \left(  \frac{\sinh2p}%
{2p}\right)  ^{d-1}\,2^{d}d\mathbf{p}\,d\mathbf{x}\label{res.int2}%
\end{equation}
where the integral (\ref{res.int2}) is an absolutely convergent complex-valued
integral. This of course is formally equivalent to (\ref{res.int}). We will
prove Theorem \ref{res.thm} at first without worrying about convergence or
other similar technicalities. Then at the end we will explain how such matters
can be dealt with.

\textit{Proof of Theorem \ref{res.thm}}. We now write the coherent states as
$\left|  \psi_{\mathbf{a}}^{\tau}\right\rangle $ to emphasize the dependence
on the dimensionless quantity $\tau=\hbar/m\omega r^{2}.$ We regard the
coherent states $\left|  \psi_{\mathbf{a}}^{\tau}\right\rangle $ as living in
some fixed ($\tau$-independent) Hilbert space (for example, $L^{2}(S^{d})$)
and given heuristically by
\begin{equation}
\left|  \psi_{\mathbf{a}}^{\tau}\right\rangle =e^{-\tau\tilde{J}^{2}/2}\left|
\delta_{\mathbf{a}}\right\rangle ,\quad\mathbf{a}\in S_{\mathbb{C}}%
^{d},\label{psia}%
\end{equation}
where $\left|  \delta_{\mathbf{a}}\right\rangle $ is a position eigenvector.
Our strategy is essentially the one proposed by T. Thiemann in a more general
setting in \cite[Sect. 2.3]{T1}. We begin with two lemmas that allow us to
carry out this strategy explicitly in this situation. The proofs of these
lemmas are given at the end of the proof of Theorem \ref{res.thm}.

\begin{lemma}
\label{volume.lem}The measure
\[
\left(  \frac{\sinh2p}{2p}\right)  ^{d-1}\,2^{d}d\mathbf{p}\,d\mathbf{x}%
\]
is invariant under the action of $\mathrm{SO}(d+1;\mathbb{C})$ on
$S_{\mathbb{C}}^{d}\cong T^{\ast}(S^{d}).$
\end{lemma}

\begin{lemma}
\label{laplace.lem}Let $J_{\mathbf{a}}^{2}$ and $J_{\mathbf{\bar{a}}}^{2}$
denote the differential operators on $S_{\mathbb{C}}^{d}$ given by
\begin{align*}
J_{\mathbf{a}}^{2}=-\sum_{k<l}\left(  a_{l}\frac{\partial}{\partial a_{k}%
}-a_{k}\frac{\partial}{\partial a_{l}}\right)  ^{2}\\
J_{\mathbf{\bar{a}}}^{2}=-\sum_{k<l}\left(  \bar{a}_{l}\frac{\partial
}{\partial\bar{a}_{k}}-\bar{a}_{k}\frac{\partial}{\partial\bar{a}_{l}}\right)
^{2}.
\end{align*}
Let $\phi$ be a smooth, even, real-valued function on $\mathbb{R}$ and
consider the function on $S_{\mathbb{C}}^{d}$ given by
\[
\phi\left(  2p\right)
\]
where $p$ is regarded as a function of $\mathbf{a}$ by means of
(\ref{a.explicit}). Then
\begin{align*}
J_{\mathbf{a}}^{2}\phi\left(  2p\right)  =J_{\mathbf{\bar{a}}}^{2}\phi\left(
2p\right) \\
=-\left[  \frac{\partial^{2}\phi}{\partial R^{2}}+(d-1)\frac{\cosh R}{\sinh
R}\frac{\partial\phi}{\partial R}\right]  _{R=2p}.
\end{align*}
\end{lemma}

Assuming for now the two lemmas, we proceed with the proof of the resolution
of the identity. Because the coherent states depend holomorphically on
$\mathbf{a}$ they satisfy
\[
J_{\mathbf{\bar{a}}}^{2}\left|  \psi_{\mathbf{a}}^{\tau}\right\rangle =0.
\]
Furthermore, it follows from the definition of the coherent states that
\[
\frac{d}{d\tau}\left|  \psi_{\mathbf{a}}^{\tau}\right\rangle =\frac{1}%
{2}J_{\mathbf{a}}^{2}\left|  \psi_{\mathbf{a}}^{\tau}\right\rangle
\]
The proof of this is essentially the standard calculation of the action of
$\tilde{J}^{2}$ in the position representation. It then follows that the
projection operator $\left|  \psi_{\mathbf{a}}^{\tau}\right\rangle
\left\langle \psi_{\mathbf{a}}^{\tau}\right|  $ satisfies the operator-valued
differential equation
\[
\frac{\partial}{\partial\tau}\left|  \psi_{\mathbf{a}}^{\tau}\right\rangle
\left\langle \psi_{\mathbf{a}}^{\tau}\right|  =\frac{1}{2}\left(
J_{\mathbf{a}}^{2}+J_{\mathbf{\bar{a}}}^{2}\right)  \left|  \psi_{\mathbf{a}%
}^{\tau}\right\rangle \left\langle \psi_{\mathbf{a}}^{\tau}\right|  .
\]

Now we let
\begin{equation}
\beta\left(  p\right)  =2^{d}\left(  \frac{\sinh2p}{2p}\right)  ^{d-1}%
.\label{beta.form}%
\end{equation}
Since by Lemma \ref{volume.lem} the measure $\beta\left(  p\right)
\,d\mathbf{p}\,d\mathbf{x}$ is invariant under the action of $\mathrm{SO}%
(d+1;\mathbb{C}$) the operators $J_{\mathbf{a}}^{2}$ and $J_{\mathbf{\bar{a}}%
}^{2} $ are self-adjoint in $L^{2}(S_{\mathbb{C}}^{d},\beta(p)\,d\mathbf{p}%
\,d\mathbf{x}).$ Thus differentiating under the integral sign and using the
self-adjointness gives
\begin{align*}
& \frac{d}{d\tau}\int_{\mathbf{x}\in S^{d}}\int_{\mathbf{p}\cdot\mathbf{x}%
=0}\left|  \psi_{\mathbf{a}}^{\tau}\right\rangle \left\langle \psi
_{\mathbf{a}}^{\tau}\right|  \,\,\nu\left(  2\tau,2p\right)  \beta
(p)\,d\mathbf{p}\,d\mathbf{x}\\
& =\int_{\mathbf{x}\in S^{d}}\int_{\mathbf{x}\cdot\mathbf{p}=0}\left|
\psi_{\mathbf{a}}^{\tau}\right\rangle \left\langle \psi_{\mathbf{a}}^{\tau
}\right|  \,\left[  \frac{\partial}{\partial\tau}+\frac{1}{2}\left(
J_{\mathbf{a}}^{2}+J_{\mathbf{\bar{a}}}^{2}\right)  \nu\left(  2\tau
,2p\right)  \right]  \beta(p)\,d\mathbf{p}\,d\mathbf{x}.
\end{align*}

Lemma \ref{laplace.lem} and the differential equation satisfied by $\nu(s,R) $
tell us that the last integral is zero. Thus the operator on the right in
(\ref{res.int}) is independent of $\tau.$ On the other hand, the initial
conditions for $\nu$ imply that as $\tau$ tends to zero the measure
$\nu\left(  2\tau,2p\right)  \beta(p)\,d\mathbf{p}\,d\mathbf{x}$ ``collapses''
to the Lebesgue measure $d\mathbf{x}$ on the real sphere, i.e. the set where
$\mathbf{p}=0.$ Furthermore, if we consider coherent states $\left|
\psi_{\mathbf{x},\mathbf{p}}^{\tau}\right\rangle $ with $\mathbf{p}=0,$ these
become simply $\left|  \delta_{\mathbf{x}}\right\rangle $ in the
$\tau\downarrow0$ limit. Thus
\[
\lim_{\tau\downarrow0}\int_{\mathbf{x}\in S^{d}}\int_{\mathbf{p}%
\cdot\mathbf{x}=0}\left|  \psi_{\mathbf{x},\mathbf{p}}^{\tau}\right\rangle
\left\langle \psi_{\mathbf{x},\mathbf{p}}^{\tau}\right|  \,\nu\left(
2\tau,2p\right)  \beta(p)\,d\mathbf{p}\,d\mathbf{x}=\int_{S^{d}}\left|
\delta_{\mathbf{x}}\right\rangle \left\langle \delta_{\mathbf{x}}\right|
\,d\mathbf{x}=I.
\]
Since the value of the first integral is independent of $\tau$ this shows that
the integral equals the identity for all $\tau.$ $\square$

It remains now to prove Lemmas \ref{volume.lem} and \ref{laplace.lem}. We
begin with the second lemma.

\textit{Proof of Lemma \ref{laplace.lem}}. Note that expressions such as
$\partial/\partial a_{k}$ do not make sense when applied to a function that is
defined only on the complex sphere $S_{\mathbb{C}}^{d}.$ So the operator
$a_{l}\partial/\partial a_{k}-a_{k}\partial/\partial a_{l}$ (and its complex
conjugate) should be interpreted as follows. Given a smooth function $f$ on
$S_{\mathbb{C}}^{d},$ extend $f$ smoothly to a neighborhood of $S_{\mathbb{C}%
}^{d},$ then apply $a_{l}\partial/\partial a_{k}-a_{k}\partial/\partial
a_{l},$ and then restrict again to $S_{\mathbb{C}}^{d}.$ Since $(a_{l}%
\partial/\partial a_{k}-a_{k}\partial/\partial a_{l})a^{2}=0$ the derivatives
are all in directions tangent to $S_{\mathbb{C}}^{d}.$ This means that the
value of the operator on $S_{\mathbb{C}}^{d}$ is independent of the choice of
the extension. It is in this way that $J_{\mathbf{a}}^{2}$ and
$J_{\mathbf{\bar{a}}}^{2}$ are to be interpreted as operators on
$S_{\mathbb{C}}^{d}.$

Now let $R=2p$ and let $\alpha=\left|  \mathbf{a}\right|  ^{2}=\Sigma\left|
a_{k}\right|  ^{2}.$ Then (\ref{a.explicit}) (with $r=m\omega=1$) tells us
that
\[
\alpha:=\left|  \mathbf{a}\right|  ^{2}=\cosh^{2}p+\sinh^{2}p=\cosh2p.
\]
So $R=2p=\cosh^{-1}\alpha.$ We now need to apply $J_{\mathbf{a}}^{2}$ to a
function of the form $\phi(R),$ which we do by using the chain rule
\[
\frac{\partial\phi}{\partial a_{k}}=\frac{d\phi}{dR}\frac{dR}{d\alpha}%
\frac{\partial\alpha}{\partial a_{k}}.
\]

Calculation shows that (for $k\neq l$)
\begin{align}
\left(  a_{l}\frac{\partial}{\partial a_{k}}-a_{k}\frac{\partial}{\partial
a_{l}}\right)  ^{2}\phi(R)  & =\frac{\left(  a_{k}\bar{a}_{l}-a_{l}\bar{a}%
_{k}\right)  ^{2}}{\left|  \mathbf{a}\right|  ^{4}-1}\frac{\partial^{2}\phi
}{\partial R^{2}}\nonumber\\
& -\frac{\left(  \left|  a_{k}\right|  ^{2}+\left|  a_{l}\right|  ^{2}\right)
\left(  \left|  \mathbf{a}\right|  ^{4}-1\right)  +\left|  \mathbf{a}\right|
^{2}\left(  a_{k}\bar{a}_{l}-a_{l}\bar{a}_{k}\right)  ^{2}}{\left(  \left|
\mathbf{a}\right|  ^{4}-1\right)  ^{3/2}}\frac{\partial\phi}{\partial
R}.\label{first.ja}%
\end{align}
We now note that
\begin{align}
\sum_{k<l}\left(  \left|  a_{k}\right|  ^{2}+\left|  a_{l}\right|
^{2}\right)   & =\frac{1}{2}\sum_{k,l}\left(  1-\delta_{kl}\right)  \left(
\left|  a_{k}\right|  ^{2}+\left|  a_{l}\right|  ^{2}\right) \nonumber\\
& =\frac{1}{2}\left[  2(d+1)\left|  \mathbf{a}\right|  ^{2}-2\left|
\mathbf{a}\right|  ^{2}\right]  =d\,\left|  \mathbf{a}\right|  ^{2}%
.\label{sum.id}%
\end{align}
We use also the easily verified identity
\begin{align}
\sum_{k<l}\left(  a_{k}\bar{a}_{l}-a_{l}\bar{a}_{k}\right)  ^{2}  & =-\left(
\left|  \mathbf{a}\right|  ^{4}-\left|  a^{2}\right|  ^{2}\right) \nonumber\\
& =-\left(  \left|  \mathbf{a}\right|  ^{4}-1\right)  ,\label{magic.id}%
\end{align}
where the first line is valid everywhere and the second line is valid on the
complex sphere $S_{\mathbb{C}}^{d}=\left\{  a^{2}=1\right\}  .$

Using (\ref{sum.id}) and (\ref{magic.id}) we get, upon summing (\ref{first.ja}%
) over $k<l,$
\[
\sum_{k<l}\left(  a_{l}\frac{\partial}{\partial a_{k}}-a_{k}\frac{\partial
}{\partial a_{l}}\right)  ^{2}\phi(R)=-\frac{\partial^{2}\phi}{\partial R^{2}%
}-\left(  d-1\right)  \frac{\left|  \mathbf{a}\right|  ^{2}}{\sqrt{\left|
\mathbf{a}\right|  ^{4}-1}}\frac{\partial\phi}{\partial R}.
\]
Recalling that $\left|  \mathbf{a}\right|  ^{2}=\cosh R,$ so that
$\sqrt{\left|  \mathbf{a}\right|  ^{4}-1}=\sinh R,$ we get the claimed
formula. This completes the proof of the second lemma (with the same argument
for the conjugated case). $\square$

\textit{Proof of Lemma \ref{volume.lem}}. Our proof is indirect and uses Lemma
\ref{laplace.lem}. We regard $S_{\mathbb{C}}^{d}$ as the quotient
$\mathrm{SO}(d+1;\mathbb{C})/\mathrm{SO}(d;\mathbb{C}).$ Since both
$\mathrm{SO}(d+1;\mathbb{C})$ and $\mathrm{SO}(d;\mathbb{C})$ are unimodular,
general principles \cite[Thm. 8.36]{Kn} tell us that there is a smooth
$\mathrm{SO}(d+1;\mathbb{C})$-invariant measure on $S_{\mathbb{C}}^{d}$ and
that it is unique up to a constant. This measure must be in particular
$\mathrm{SO}(d+1) $-invariant, which means that it must be of the form
$\gamma(p)\,d\mathbf{p}\,d\mathbf{x},$ since $d\mathbf{p}\,d\mathbf{x}$ is
also $\mathrm{SO}(d+1)$-invariant. Now the operator $J_{\mathbf{a}}^{2}$ must
be self-adjoint with respect to the $\mathrm{SO}(d+1;\mathbb{C})$-invariant
measure. In particular, $J_{\mathbf{a}}^{2}$ must be self-adjoint when
restricted to the space of $\mathrm{SO}(d+1)$-invariant functions, which can
all be written in the form $f(\mathbf{a})=\phi(2p),$ as in Lemma
\ref{laplace.lem}.

Meanwhile, according to Lemma \ref{laplace.lem}, on $\mathrm{SO}%
(d+1)$-invariant functions $J_{\mathbf{a}}^{2}$ is just the hyperbolic
Laplacian, re-scaled by a factor of 2. This operator is therefore self-adjoint
(on $\mathrm{SO}(d+1)$-invariant functions) with respect to the measure
$\beta(p)\,d\mathbf{p}\,d\mathbf{x},$ which is just hyperbolic volume measure
with the same re-scaling.

We conclude, then, that on $\mathrm{SO}(d+1)$-invariant functions,
$J_{\mathbf{a}}^{2}$ is self-adjoint with respect to both the measures
$\gamma(p)\,d\mathbf{p}\,d\mathbf{x}$ and $\beta(p)\,d\mathbf{p}%
\,d\mathbf{x}.$ From this it follows that
\begin{equation}
\left[  \frac{\partial^{2}g}{\partial R^{2}}+(d-1)\frac{\cosh R}{\sinh R}%
\frac{\partial g}{\partial R}\right]  _{R=2p}=0\label{g.eq}%
\end{equation}
where $g(p)=\gamma(p)/\beta(p).$ But since both $\gamma$ and $\beta$ are
smooth, $\mathrm{SO}(d+1)$-invariant functions on $S_{\mathbb{C}}^{d}$ we must
have $\partial g/\partial R|_{R=0}=0.$ Solving (\ref{g.eq}) gives $\partial
g/\partial R=c\exp\left[  -(d-1)\int\coth S\,\,dS\right]  ,$ so we have
\[
\left.  \frac{\partial g}{\partial R}\right|  _{R=0}=c\lim_{\varepsilon
\rightarrow0}\exp\left[  (d-1)\int_{\varepsilon}^{1}\coth S\,\,dS\right]  =0,
\]
which can occur only if $c=0,$ i.e. if $g$ is constant. Thus $\gamma$ is a
constant multiple of $\beta,$ which is what we want to show. $\square$

We conclude this section with a few remarks about technicalities in the proof
of the resolution of the identity. We have already said that the integral in
Theorem \ref{res.thm} is to be interpreted in the weak sense, as in
(\ref{res.int2}). We first establish (\ref{res.int2}) in the case where
$\left|  \phi_{1}\right\rangle $ and $\left|  \phi_{2}\right\rangle $ are
finite linear combinations of spherical harmonics. In that case it can be
shown that the integrand $\left\langle \phi_{1}|\psi_{\mathbf{x},\mathbf{p}%
}^{\tau}\right\rangle \left\langle \psi_{\mathbf{x},\mathbf{p}}^{\tau}%
|\phi_{2}\right\rangle $ grows only exponentially with $\mathbf{p}.$ Since
$\nu$ has faster-than-exponential decay (namely, Gaussian) the integral
(\ref{res.int2}) is convergent. In this case there is not much difficulty in
justifying the formal manipulations we have made, such as interchanging
derivatives with the integral and integrating by parts. Then once
(\ref{res.int2}) is established for such ``nice'' vectors, a simple passage to
the limit will establish it for all $\left|  \phi_{1}\right\rangle $ and
$\left|  \phi_{2}\right\rangle $ in the Hilbert space. See \cite{H1} or
\cite{St} for more details on these technicalities.

\section{The Segal--Bargmann representation\label{sb.sec}}

As shown in Section \ref{euclid.sec}, any two irreducible unitary
representations of $\tilde{E}(d+1)$ satisfying (\ref{qconstraint.1}) and
(\ref{qconstraint.2}) are equivalent. The simplest concrete realization of
such representations is the position representation, in which the Hilbert
space is $L^{2}(S^{d}),$ the position operators act by multiplication, and the
angular momentum operators act as differential operators given by
\begin{equation}
J_{kl}=-i\hbar\left(  x_{l}\frac{\partial}{\partial x_{k}}-x_{k}\frac
{\partial}{\partial x_{l}}\right)  .\label{j.form}%
\end{equation}

The resolution of the identity for the coherent states can be used to give
another realization, the (generalized) Segal--Bargmann representation. In the
Segal--Bargmann representation the Hilbert space is the space of holomorphic
functions on $S_{\mathbb{C}}^{d}$ that are square-integrable with respect to
the density occurring in the resolution of the identity. In this
representation the action of the position operators $X_{k}$ is somewhat
complicated, but the action of the creation operators (the adjoints of the
annihilation operators) becomes simply multiplication by $a_{k}.$ The
resolution of the identity can be re-interpreted as the unitary intertwining
map between these two representations, that is, the generalized
Segal--Bargmann transform.

Specifically, given any function $f$ in $L^{2}(S^{d})$ we define the
Segal--Bargmann transform $Cf$ of $f$ by
\begin{equation}
Cf\left(  \mathbf{a}\right)  =\left\langle \psi_{\mathbf{\bar{a}}%
}|f\right\rangle \label{c.def}%
\end{equation}
Then for any $f,$ $Cf(\mathbf{a})$ is a holomorphic function of $\mathbf{a}\in
S_{\mathbb{C}}^{d}.$ Note that in the interests of consistency with
\cite{H1,St} we have put a complex conjugate on the $\mathbf{a}$ in the
(\ref{c.def}), so that the dependence of $Cf$ on $\mathbf{a}\in S_{\mathbb{C}%
}^{d} $ is holomorphic rather than anti-holomorphic. The Segal--Bargmann
transform can be computed as
\begin{equation}
Cf(\mathbf{a})=\int_{S^{d}}\rho_{\tau}(\mathbf{a},\mathbf{x})f(\mathbf{x}%
)\,d\mathbf{x},\quad\mathbf{a}\in S_{\mathbb{C}}^{d}.\label{c.def2}%
\end{equation}
Here $\rho_{\tau}(\mathbf{a},\mathbf{x})$ is the heat kernel on $S^{d},$ with
the $\mathbf{a}$ variable extended by analytic continuation from $S^{d}$ to
$S_{\mathbb{C}}^{d}.$

\begin{theorem}
[Segal--Bargmann transform]\label{sb.thm}The map $C$ defined by (\ref{c.def})
or (\ref{c.def2}) is a unitary map of $L^{2}(S^{d},d\mathbf{x})$ onto
$\mathcal{H}L^{2}(S_{\mathbb{C}}^{d},\nu),$ where $\mathcal{H}L^{2}%
(S_{\mathbb{C}}^{d},\nu)$ denotes the space of holomorphic functions $F$ on
$S_{\mathbb{C}}^{d}$ for which
\[
\int_{\mathbf{x}\in S^{d}}\int_{\mathbf{p}\cdot\mathbf{x}=0}\left|
F(\mathbf{a}\left(  \mathbf{x},\mathbf{p}\right)  )\right|  ^{2}\nu\left(
2\tau,2p\right)  \left(  \frac{\sinh2p}{2p}\right)  ^{d-1}\,2^{d}%
d\mathbf{p}\,d\mathbf{x}<\infty.
\]
\end{theorem}

The isometricity of the map $C$ as a map from $L^{2}(S^{d})$ into
$L^{2}(S_{\mathbb{C}}^{d},\nu)$ is equivalent to the resolution of the
identity. (Compare (\ref{res.int2}).) That $C$ maps into the holomorphic
subspace of $L^{2}(S_{\mathbb{C}}^{d},\nu)$ follows from the holomorphic
dependence of the coherent states on $\mathbf{a}.$ It remains only to show
that the image of $C$ is \textit{all} of $\mathcal{H}L^{2}(S_{\mathbb{C}}%
^{d},\nu).$ The proof of this is a fairly straightforward density argument
using spherical harmonics, which we omit. (See Section 4 of \cite{St} and
Section 8 of \cite{H1}.)

In the Segal--Bargmann space $\mathcal{H}L^{2}(S_{\mathbb{C}}^{d},\nu)$ the
angular momentum operators act by the holomorphic analog of (\ref{j.form}),
namely,
\[
J_{kl}=-i\hbar\left(  a_{l}\frac{\partial}{\partial a_{k}}-a_{k}\frac
{\partial}{\partial a_{l}}\right)  .
\]
Meanwhile, the \textit{creation} operators, defined as the adjoints of the
annihilation operators, are given by
\[
A_{k}^{\dagger}F(\mathbf{a})=a_{k}F(\mathbf{a}).
\]
The annihilation operators can be described as \textit{Toeplitz operators}.
This means that
\[
A_{k}F=P(\bar{a}_{k}F),
\]
where $P$ is the orthogonal projection from the full $L^{2}$-space
$L^{2}(S_{\mathbb{C}}^{d},\nu)$ onto the holomorphic subspace. (See
\cite{H6}.) In the Segal--Bargmann representation the action of the position
operators is more complicated.

Another important feature of the Segal--Bargmann representation is the
reproducing kernel identity. Recall from Section \ref{coherent.sec} that the
reproducing kernel $R_{\tau}(\mathbf{a,b})=\left\langle \left.  \psi
_{\mathbf{b}}\right|  \psi_{\mathbf{a}}\right\rangle $ is holomorphic in
$\mathbf{a}$ and anti-holomorphic in $\mathbf{b}.$ We then have the following
result, which follows easily from general principles. (See, for example,
Section 9 of \cite{H1}.)

\begin{proposition}
[Reproducing kernel identity]For any $F\in\mathcal{H}L^{2}(S_{\mathbb{C}}%
^{d},\nu)$ we have
\[
F(\mathbf{a})=\int_{\mathbf{x}\in S^{d}}\int_{\mathbf{p}\cdot\mathbf{x}%
=0}R_{\tau}(\mathbf{a},\mathbf{b})F(\mathbf{b}\left(  \mathbf{x}%
,\mathbf{p}\right)  )\nu\left(  2\tau,2p\right)  \left(  \frac{\sinh2p}%
{2p}\right)  ^{d-1}\,2^{d}d\mathbf{p}\,d\mathbf{x}.
\]
Here $R_{\tau}(\mathbf{a},\mathbf{b})=\rho_{2\tau}(\mathbf{a},\mathbf{\bar{b}%
})$ is the reproducing kernel, and the integral is absolutely convergent.
\end{proposition}

The Segal--Bargmann representation can be thought of as defining a
\textit{phase space wave function} for a quantum particle on the sphere, which
is related to the position wave function by the Segal--Bargmann transform.
This phase space wave function can then be turned into a \textit{phase space
probability density} in the usual way: if $f$ is a unit vector in $L^{2}%
(S^{d})$ then the associated probability density is
\begin{equation}
\left|  Cf(\mathbf{a}\left(  \mathbf{x},\mathbf{p}\right)  )\right|  ^{2}%
\nu(2\tau,2p)\,\left(  \frac{\sinh2p}{2p}\right)  ^{d-1}2^{d}.\label{husimi}%
\end{equation}
This is a manifestly non-negative function on the phase space that integrates
to one. In the $\mathbb{R}^{d}$ case \cite{H6}, the expression corresponds to
the \textit{Husimi function} of $f.$

If one takes the probability density (\ref{husimi}) and integrates out the
momentum variables one will \textit{not} get the standard position probability
density $\left|  f(\mathbf{x})\right|  ^{2}$ (even in the $\mathbb{R}^{d}$
case). That is, with this definition, the position probability density cannot
be obtained from the phase space probability density by integrating out the
momentum variables. On the other hand, there is a nice inversion formula for
the generalized Segal--Bargmann transform that can be stated roughly as
follows: the position \textit{wave function} can be obtained from the phase
space \textit{wave function} by integrating out the momentum variables.

\begin{theorem}
[Inversion Formula]\label{inv.thm}Given any function $f$ in $L^{2}(S^{d}),$
let $F=Cf$ be the Segal--Bargmann transform of $f.$ Then $f$ may be recovered
from $F$ by the formula
\begin{equation}
f(\mathbf{x})=\int_{\mathbf{p}\cdot\mathbf{x}=0}F(\mathbf{a}\left(
\mathbf{x},\mathbf{p}\right)  )\nu(\tau,p)\left(  \frac{\sinh p}{p}\right)
^{d-1}\,d\mathbf{p}.\label{inverse.int}%
\end{equation}
\end{theorem}

This result is a special case of \cite{St}; the group analog of this inversion
formula was given in \cite{H2}. An analog of this formula holds also in the
$\mathbb{R}^{d}$ case \cite[Sect. 4]{H4}, but does not seem to be well known.
Note that whereas the resolution of the identity involves $\nu(2\tau,2p),$ the
inversion formula involves $\nu(\tau,p)$.

This statement of the inversion formula is a bit imprecise, because we have
glossed over the question of the convergence of the integral in
(\ref{inverse.int}). The integral cannot always be convergent, since a general
function $f$ in $L^{2}(S^{d})$ can have singularities. As shown in Theorems 1
and 2 of \cite{St}, we have the following two precise statements. First, if
$f$ is sufficiently smooth, then the integral in (\ref{inverse.int}) is
absolutely convergent for all $\mathbf{x}$ and is equal to $f(\mathbf{x}).$
Second, for any $f\in L^{2}(S^{d})$ we have
\[
f(\mathbf{x})=\lim_{R\rightarrow\infty}\int_{\substack{\mathbf{p}%
\cdot\mathbf{x}=0 \\p\leq R}}F(\mathbf{a}\left(  \mathbf{x},\mathbf{p}\right)
)\nu(\tau,p)\left(  \frac{\sinh p}{p}\right)  ^{d-1}\,d\mathbf{p}%
\]
where the limit is in the topology of $L^{2}(S^{d}).$

We will describe the proof of Theorem \ref{inv.thm} in greater detail in the
setting of general compact symmetric spaces. Here we give only the barest
outline. The Cauchy--Riemann equations on $S_{\mathbb{C}}^{d}$ imply that,
when applied to holomorphic functions, the hyperbolic Laplacian in the
momentum variables is the negative of the spherical Laplacian in the position
variables, just as for holomorphic functions on $\mathbb{C}$ we have
$\partial^{2}F/\partial y^{2}=-\partial^{2}F/\partial x^{2}$. For this result
to hold, we must omit the re-scaling of the momentum variables that is present
in the resolution of the identity; hence the inversion formula involves
$\nu(\tau,p)$ rather than $\nu(2\tau,2p).$ The integration in
(\ref{inverse.int}) against the hyperbolic heat kernel is computing the
forward heat equation in the momentum variables, which for holomorphic
functions is then the same as the backward heat equation in the position
variables. Since the Segal--Bargmann transform may be computed in terms of the
forward heat equation in the position variables, (\ref{inverse.int}) is
inverting the Segal--Bargmann transform. Although this is the basic idea of
the proof, the convergence questions are more subtle.

Note that there are, besides (\ref{inverse.int}), many other inversion
formulas for the Segal--Bargmann transform. The ``overcompleteness'' of the
coherent states means that there is a lot of redundant information in the
Segal--Bargmann transform, and therefore many different ways that one can
recover $f$ from $Cf.$ To look at it another way, it is possible to have many
different integrals that all give the same value when applied to holomorphic
functions, as in the Cauchy integral formula. Of particular importance is the
inversion formula
\[
f(\mathbf{x}^{\prime})=\int_{S_{\mathbb{C}}^{d}}\overline{\rho_{\tau
}(\mathbf{a},\mathbf{x}^{\prime})}F(\mathbf{a}\left(  \mathbf{x}%
,\mathbf{p}\right)  )\,\nu(2\tau,2p)\,\left(  \frac{\sinh2p}{2p}\right)
^{d-1}\,2^{d}d\mathbf{p}\,d\mathbf{x},
\]
where $\rho_{\tau}$ is the analytical continuation of the heat kernel for
$S^{d}.$ This formula is obtained by noting that $C$ is isometric, and
therefore its inverse is its adjoint. One can apply the above integral to any
function $F$ in $L^{2}(S_{\mathbb{C}}^{d},\nu)$ (not necessarily holomorphic),
in which case we have $f=C^{-1}PF,$ where $PF$ is the orthogonal projection of
$F$ onto the holomorphic subspace of $L^{2}(S_{\mathbb{C}}^{d},\nu).$ See
\cite[Sect. 9]{H1} and \cite[Eq. (6.13)]{KR2}.

\section{The $\mathbb{R}^{d}$ case\label{rd.sec}}

We verify in this section that the methods in this paper, when applied to the
$\mathbb{R}^{d}$ case, do indeed reproduce the canonical coherent states. Our
``complexifier'' is $1/\omega$ times the kinetic energy function, namely,
$p^{2}/2m\omega.$ (In the $\mathbb{R}^{d}$ case the kinetic energy cannot be
expressed in terms of the angular momentum.) Then we define
\begin{align*}
a_{k}  & =e^{i\left\{  \cdot,\text{ complexifier}\right\}  }x_{k}\\
& =\sum_{n=0}^{\infty}\left(  \frac{i}{2m\omega}\right)  ^{n}\frac{1}%
{n!}\underset{n}{\underbrace{\left\{  \cdots\left\{  \left\{  x_{k}%
,p^{2}\right\}  ,p^{2}\right\}  ,\cdots,p^{2}\right\}  }}.
\end{align*}
Since $\left\{  x_{k},p^{2}\right\}  =2p_{k}$ and $\left\{  \left\{
x_{k},p^{2}\right\}  ,p^{2}\right\}  =0$ we obtain
\[
a_{k}=x_{k}+i\frac{p_{k}}{m\omega}.
\]
This is, up to an overall constant, the standard complex coordinate on phase
space. More generally one can apply the same method to any function of the
$x_{k}$'s, and one will obtain the corresponding function of $a_{k}.$ For
example, it is easily verified by induction that
\[
e^{i\left\{  \cdot,\text{ complexifier}\right\}  }\left(  x_{k}^{n}\right)
=\left(  x_{k}+i\frac{p_{k}}{m\omega}\right)  ^{n}%
\]
for all positive integers $n.$

Similarly on the quantum side if we define the complexifier to be
$P^{2}/2m\omega$ and
\begin{align*}
A_{k}  & =e^{i[\cdot,\text{complexifier}]/i\hbar}X_{k}\\
& =\sum_{n=0}^{\infty}\frac{1}{(2m\omega\hbar)^{n}}\frac{1}{n!}\left[
\cdots\left[  \left[  X_{k},P^{2}\right]  ,P^{2}\right]  \cdots,P^{2}\right]
\end{align*}
we get simply
\[
A_{k}=X_{k}+i\frac{P_{k}}{m\omega}.
\]
This is, up to an overall constant, the usual annihilation operator. Applying
the same procedure to any function of the $X_{k}$'s will give the
corresponding function of the $A_{k}$'s.

Following the same normalization procedure as in the sphere case we obtain
coherent states given by
\[
\left|  \psi_{\mathbf{a}}\right\rangle =e^{-P^{2}/2m\omega\hbar}\left|
\delta_{\mathbf{a}}\right\rangle ,
\]
at first for $\mathbf{a}\in\mathbb{R}^{d}$ and then by analytic continuation
for any $\mathbf{a}\in\mathbb{C}^{d}$. In the $\mathbb{R}^{d}$ case we have
the formula
\[
\left|  \psi_{\mathbf{a}}\right\rangle =e^{i\mathbf{a}\cdot\mathbf{P}/\hbar
}\left|  \psi_{0}\right\rangle .
\]
This normalization coincides with what Hecht \cite{Hec} calls Type I coherent
states. In the position representation we have
\[
\left\langle \left.  \delta_{\mathbf{x}}\right|  \psi_{\mathbf{a}%
}\right\rangle =\left(  2\pi\hbar/m\omega\right)  ^{-d/2}\exp\left[
-\frac{\left(  \mathbf{x}-\mathbf{a}\right)  ^{2}}{2\hbar/m\omega}\right]  .
\]

With this normalization of the coherent states the resolution of the identity
takes the form
\[
I=\int_{\mathbb{C}^{d}}\left|  \psi_{\mathbf{a}}\right\rangle \left\langle
\psi_{\mathbf{a}}\right|  \,\gamma\left(  \mathbf{a}\right)  \,d\mathbf{a},
\]
where $d\mathbf{a}$ is $2d$-dimensional Lebesgue measure and where $\gamma$ is
the density
\[
\gamma\left(  \mathbf{a}\right)  =\left(  \frac{\pi\hbar}{m\omega}\right)
^{-d/2}\exp\left[  -\frac{\left(  \operatorname{Im}\mathbf{a}\right)  ^{2}%
}{\hbar/m\omega}\right]  .
\]
The associated Segal--Bargmann space is the space of holomorphic functions on
$\mathbb{C}^{d}$ that are square-integrable with respect to the density
$\gamma.$ This normalization of the Segal--Bargmann space is different from
that of Segal \cite{Se} and Bargmann \cite{B}, because of the different
normalization of the coherent states. See \cite[Sect. 6]{H6} for comparisons
with the conventions of Segal and of Bargmann.

To compare this to what we have in the sphere case, let $\sigma=\hbar/m\omega$
and consider the Euclidean heat kernel in the imaginary directions, given by
\[
\nu(\sigma,\mathbf{a})=\left(  2\pi\sigma\right)  ^{-d/2}\exp\left[
-\frac{(\operatorname{Im}\mathbf{a})^{2}}{2\sigma}\right]  .
\]
Then $\gamma(\mathbf{a})=2^{d}\nu(2\sigma,2\mathbf{a}),$ similar to what we
have in the sphere case. Note that in the Euclidean case $\nu(2\sigma
,2\mathbf{a})$ is the same, up to an overall constant, as $\nu(\sigma
/2,\mathbf{a}).$ Thus it is hard to see the ``correct'' scaling of the space
and time variables from the Euclidean case.

An inversion formula similar to Theorem \ref{inv.thm} holds in the
$\mathbb{R}^{d}$ case; see \cite[Sect. 4]{H4}.

\section{Representation theory of the Euclidean group\label{euclid.sec}}

We consider representations by self-adjoint operators of the commutation
relations (\ref{qxj}) for the Lie algebra $\mathrm{e}(d+1).$ We further assume
that these operators are the Lie algebra representation associated to a
representation of the corresponding connected, simply connected Lie group
$\mathrm{\tilde{E}}(d+1).$ It is known that all the irreducible unitary
representations of $\mathrm{\tilde{E}}(d+1)$ can be realized in spaces of
sections of smooth vector bundles with the Lie algebra acting by smooth
differential operators. The action of the Lie algebra then extends to an
action on distributional sections, including the generalized eigenvectors of
the position operators. With this discussion in mind we will make free use of
position eigenvectors in what follows.

We apply the Wigner--Mackey method and consider an orbit of $\mathrm{Spin}%
(d+1)$ in $\mathbb{R}^{d+1},$ namely, a sphere of radius $r.$ We consider only
the case $r>0,$ in which case the little group is $\mathrm{Spin}(d).$ Fixing a
value for $r$ amounts to assuming that the operators $X_{k}$ satisfy $\sum
X_{k}^{2}=r^{2}.$

The purpose of this section is to show that the little group acts trivially if
and if the following relation holds for all $k$ and $l$:
\begin{equation}
X^{2}J_{kl}=J_{km}X_{m}X_{l}-J_{lm}X_{m}X_{k}\label{q.constr}%
\end{equation}
(sum convention). This is equivalent to the relation
\begin{equation}
J_{kl}=P_{k}X_{l}-P_{l}X_{k}\label{qjpx}%
\end{equation}
where by definition $P_{k}=r^{-2}J_{kl}X_{l}.$

Note that (\ref{q.constr}) is the quantum counterpart of the constraint to the
sphere (\ref{constraint.2}) and therefore representations of $\mathrm{\tilde
{E}}(d+1)$ satisfying it are closest to the classical motion on a sphere.
Nevertheless, other representations are of interest, and describe a quantum
particle on a sphere with internal degrees of freedom. We will consider the
general case in a future work.

Suppose now that (\ref{q.constr}) holds. We wish to show that this implies
that the representation of the little group is trivial. So we consider the
space of generalized eigenvectors for the operators $X_{k}$ satisfying
\begin{align}
X_{k}\left|  \psi\right\rangle  & =0,\quad k=1,\cdots,d\nonumber\\
X_{d+1}\left|  \psi\right\rangle  & =r.\label{x.evector}%
\end{align}
This is the space on which the little group acts, where the Lie algebra of the
little group is given by the operators $J_{kl}$ with $1\leq k,l\leq d.$ But
now if (\ref{q.constr}) holds then for $k,l\leq d$ we have
\[
r^{2}J_{kl}\left|  \psi\right\rangle =0
\]
since in that case $X_{k}\left|  \psi\right\rangle =X_{l}\left|
\psi\right\rangle =0.$ This shows (for $r>0$) that if (\ref{q.constr}) holds
then the little group acts trivially.

Consider now the quantity
\begin{equation}
W_{kl}:=X^{2}J_{kl}-J_{km}X_{m}X_{l}+J_{lm}X_{m}X_{k},\label{wkl.def}%
\end{equation}
which satisfies $W_{lk}=-W_{kl}.$ The condition (\ref{q.constr}) is equivalent
to $W_{kl}=0.$ Consider also the quantity
\begin{equation}
C:=\sum_{k<l}W_{kl}^{2}.\label{casimir}%
\end{equation}
As we will show below, $C$ is a Casimir, that is, an element of the universal
enveloping algebra of $\mathrm{e}(d+1).$ This implies that $C$ acts as $cI$ in
each irreducible representation. (The value of the constant $c$ is $r^{4}$
times the value of the quadratic Casimir for the little group in each
generalized eigenspace for the position operators.)

Let us now assume that the little group acts trivially and determine the value
of $c$ in this case. We may compute $c$ by applying $C$ to a position
eigenvector as in (\ref{x.evector}). That the little group acts trivially
means that $J_{kl}\left|  \psi\right\rangle =0$ for $k<l<d+1.$ Since also
$X_{k}\left|  \psi\right\rangle =X_{l}\left|  \psi\right\rangle =0$ for
$k<l<d+1$ we get
\[
C\left|  \psi\right\rangle =c\left|  \psi\right\rangle =\sum_{k}\left(
X^{2}J_{k,d+1}-J_{km}X_{m}X_{d+1}+J_{d+1,m}X_{m}X_{k}\right)  ^{2}\left|
\psi\right\rangle .
\]
But since $X_{m}\left|  \psi\right\rangle =0$ unless and $m=d+1$ (and since
$J_{d+1,d+1}=0$) we get that
\begin{align*}
\left(  X^{2}J_{k,d+1}-J_{km}X_{m}X_{d+1}+J_{d+1,m}X_{m}X_{k}\right)  \left|
\psi\right\rangle  & =\left(  X^{2}J_{k,d+1}-J_{k,d+1}X_{d+1}^{2}+0\right)
\left|  \psi\right\rangle \\
& =\left(  r^{2}J_{k,d+1}-r^{2}J_{k,d+1}\right)  \left|  \psi\right\rangle \\
& =0.
\end{align*}

This means that if the representation of the little group is trivial then the
constant $c$ must be zero, which means the element $C$ must be zero in that
representation. A calculation shows that for each $k<l$, $W_{kl}$ is
self-adjoint. Thus $C$ is a sum of squares of self-adjoint operators, and the
only way the sum can be zero is if each term is zero, that is, if
(\ref{q.constr}) holds. So if the little group acts trivially, (\ref{q.constr}%
) must hold, which is what we want to prove.

In the case $d=2$ (considered in \cite{KR1}) it is possible to verify that
\begin{equation}
C=X^{2}\left(  L\cdot X\right)  ^{2},\label{cl.x}%
\end{equation}
where $L$ is the angular momentum \textit{vector}, related to our angular
momentum \textit{matrix} by $L=(J_{32},J_{13},J_{12}).$ One can easily check
that at least this relation holds in each irreducible representation (which is
all that is really relevant) as follows. Both sides are Casimirs and so it
suffices to check (\ref{cl.x}) on the generalized eigenspace in
(\ref{x.evector}). But for $\left|  \psi\right\rangle $ in this space we
calculate that
\[
C\left|  \psi\right\rangle =X^{2}(L\cdot X)^{2}\left|  \psi\right\rangle
=X^{4}J_{12}^{2}\left|  \psi\right\rangle ,
\]
and indeed (\ref{cl.x}) holds. From (\ref{cl.x}) we see that taking $C=0$ is
equivalent in the $d=2$ case to taking $L\cdot X=0$ as in \cite{KR1}.

It remains only to show that the element $C$ in (\ref{casimir}) is a Casimir.
To do this we first compute the commutation relations of $W_{kl}$ with the
$J$'s and the $X$'s. These come out to be
\begin{align}
\frac{1}{i\hbar}\left[  X_{k},W_{lm}\right]   & =0\label{xw}\\
\frac{1}{i\hbar}\left[  J_{kl},W_{mn}\right]   & =\delta_{kn}W_{lm}%
+\delta_{lm}W_{kn}-\delta_{km}W_{l{}n}-\delta_{l{}n}W_{km}.\label{jw}%
\end{align}
Equation (\ref{jw}) is what we expect for a matrix operator--compare this to
the formula for $\left[  J_{kl},J_{mn}\right]  .$ Equation (\ref{xw}) implies
immediately that $C$ commutes with each $X_{k},$ and Equation (\ref{jw})
implies, after a short calculation, that $C$ commutes with each $J_{kl}.$

\section{Concluding remarks\label{conclusions.sec}}

We end this paper by discussing how the coherent states described here compare
to other coherent states that have been proposed for systems whose
configuration space is a sphere (or homogeneous space). As we have explained
in detail above, the coherent states introduced in \cite{KR1} are equivalent
to those in \cite{H1,St}, but were discovered independently and from a
different point of view.

Meanwhile, there are several other generalized Segal--Bargmann transforms for
spheres that have been considered. These are similar but not identical to each
other and were introduced by Bargmann and Todorov \cite{BT}, Rawnsley
\cite{Ra2}, Ii \cite{I}, Wada \cite{Wa}, Thomas and Wassell \cite{TWa}, and
Villegas \cite{V}. In most cases the transform is unitary, and this unitarity
can be re-formulated as a resolution of the identity for the associated
coherent states. These constructions all have in common that the coherent
states are labeled by points in the cotangent bundle \textit{minus the zero
section} (i.e. with the points of zero momentum removed). In these papers the
cotangent bundle minus the zero section is identified with the null quadric
$\left\{  a\in\mathbb{C}^{d+1}\left|  a^{2}=0\right.  \right\}  .$ This is to
be contrasted with the present paper, in which the full cotangent bundle of
the sphere is identified with the quadric $\left\{  a^{2}=r^{2}\right\}  $
with $r>0.$ Thus these constructions are inequivalent to the one considered in
this paper. Furthermore these constructions do not generalize to higher-rank
symmetric spaces \cite{Sz2}.

Besides these, there have been to our knowledge two other proposed
constructions of coherent states on spheres (and other homogeneous spaces).
These constructions, inequivalent to \cite{H1,St} and to each other, are those
of S. De Bi\`{e}vre \cite{De} and of C. Isham and J. Klauder \cite{IK}. Both
\cite{De} and \cite{IK} are based on extensions of the Perelomov approach, in
that their coherent states are all obtained from one fixed vector $\psi_{0}$
by the action of the Euclidean group. As explained in those papers, the
ordinary Perelomov approach is not applicable in this case, because the
irreducible representations of the Euclidean group are not square-integrable.
Non-square-integrability means that the usual Perelomov-type integral, which
should be a multiple of the identity operator, is in this case divergent.

De Bi\`{e}vre's approach to this problem is to apply to the fiducial vector
$\psi_{0}$ only a part of the Euclidean group. We describe just the simplest
case of \cite{De}. (This special case was worked out independently in a more
elementary way by Torresani \cite{To}.) Specifically, if we work in
$L^{2}(S^{d})$ then start with a basic coherent state $\psi_{0}$ such that a)
$\psi_{0}$ is invariant under rotations about the north pole $\mathbf{n}$ and
b) $\psi_{0}$ is supported in the northern half-sphere with a certain rate of
decay at the equator. One may think of $\psi_{0}$ being concentrated near the
north pole and approximating a state whose position is at the north pole and
whose momentum is zero. The other coherent states are then of the form
\[
\exp(i\mathbf{k}\cdot\mathbf{x})\psi_{0}(R^{-1}\mathbf{x})
\]
where we consider only pairs $(\mathbf{k},R)$ satisfying $\mathbf{k}\cdot
R\mathbf{n}=0.$ This last restriction is crucial. Since $\psi_{0}$ is
invariant under rotations about the north pole, the coherent states are
determined by the values of $\mathbf{k}$ and $R\mathbf{n}$ and are thus
labeled by points in the cotangent bundle of $S^{d}.$ The resolution of the
identity for these coherent states follows from the general procedure in
\cite{De} but can also be proved in this case by an elementary application of
the Plancherel formula. The condition that $\psi_{0}$ be supported in the
northern half-sphere is crucial to the proof.

It is clear that the coherent states considered in this paper are quite
different from those in \cite{De}. First, De Bi\`{e}vre's coherent states do
not depend holomorphically on the parameters. Second, each coherent state must
be supported in a half-sphere, hence cannot be real-analytic in the space
variable. Third, there does not seem to be any preferred choice for $\psi_{0}$
in \cite{De}, whereas for the coherent states considered here the only choice
one has to make is the value of the parameter $\omega.$

Meanwhile, Isham and Klauder use a different method of working around the
non-square-integrability of the irreducible representations of $\mathrm{E}%
(d+1).$ They use reducible representations, corresponding to integration over
some small range $\left[  r,r+\varepsilon\right]  $ of radii. This allows for
a family of coherent states invariant under the full Euclidean group and
allows a more general basic coherent state $\psi_{0},$ without any support
conditions. On the other hand it seems natural to get back to an irreducible
representation by letting $\varepsilon$ tend to zero, so that the particle is
constrained to a sphere with one fixed radius. Unfortunately, although the
representation itself does behave well under this limit (becoming irreducible)
the coherent states themselves do not have a limit as $\varepsilon$ tends to
zero. (See the remarks at the bottom of the first column on p.609 in
\cite{IK}.) This seems to be a drawback of this approach.

Finally, we mention that in the group case, the coherent states described in
this paper can be obtained by means of geometric quantization, as shown in
\cite{H9}. This means that in the group case the coherent states are of
``Rawnsley type'' \cite{Ra}. However, this result does not carry over to the
case of general compact symmetric spaces. In particular the results of
\cite{H9} apply only to those spheres that are also groups, namely,
$S^{1}=\mathrm{U}(1)$ and $S^{3}=\mathrm{SU}(2).$

\section{Acknowledgments}
The second author was supported in part by NSF Grant DMS-9970882.

\end{document}